\documentclass[article]{elsarticle}

\usepackage{soul}
\usepackage{rotating}
\usepackage{changepage}
\usepackage[flushleft]{threeparttable}
\usepackage{pdflscape}

\usepackage{color}
\usepackage{amsmath} 
\usepackage{lineno,hyperref}
\modulolinenumbers[5]

\begin{document}

\begin{frontmatter}

\title{Knowledge Discovery in Cryptocurrency Transactions: A Survey}


\author[cityucom]{Xiao Fan Liu\corref{mycorrespondingauthor}}
\cortext[mycorrespondingauthor]{Corresponding author}
\ead{xf.liu@cityu.edu.hk}
\author[seu]{Xin-Jian Jiang}
\author[seu]{Si-Hao Liu}
\author[cityuee]{Chi Kong Tse}

\address[cityucom]{Department of Media and Communication, City University of Hong Kong, Hong Kong SAR, China}
\address[seu]{School of Computer Science and Engineering, Southeast University, Nanjing, China}
\address[cityuee]{Department of Electrical Engineering, City University of Hong Kong, Hong Kong SAR, China}

\begin{abstract}
Cryptocurrencies gain trust in users by publicly disclosing the full creation and transaction history. In return, the transaction history faithfully records the whole spectrum of cryptocurrency user behaviors. This article analyzes and summarizes the existing research on knowledge discovery in the cryptocurrency transactions using data mining techniques. Specifically, we classify the existing research into three aspects, i.e., transaction tracings and blockchain address linking, the analyses of collective user behaviors, and the study of individual user behaviors. For each aspect, we present the problems, summarize the methodologies, and discuss major findings in the literature. Furthermore, an enumeration of transaction data parsing and visualization tools and services is also provided. Finally, we outline several future directions in this research area, such as the current rapid development of Decentralized Finance (De-Fi) and digital fiat money.
\end{abstract}


\end{frontmatter}


\setcounter{tocdepth}{2}
\tableofcontents 

\section{Introduction}
\label{sec:introduction}
As of 2020, more than 7 000 cryptocurrencies are actively trading in more than 20 000 online exchanges. Their total market capitalization has exceeded USD 300 billion \cite{website:coinmarketcap}. Although the cryptocurrencies are not backed by any tangible assets, they gain trusts from users by publicly disclosing the full creation and transaction history in peer-to-peer blockchain networks.

Each transaction in the blockchain consists of transferring a virtual value from a virtual identity, i.e., a blockchain address or a set of addresses, to another. The sizes of transaction records are quickly expanding. As of 2020, the total transaction volumes of Bitcoin and Ethereum (the top two cryptocurrencies by market capitalization) have exceeded 500 million \cite{website:blockchain.info} and 600 million \cite{website:etherscan}, respectively. These transaction histories, although technically challenging in extracting, transforming, and analyzing, have given us an unprecedented opportunity to study the panorama of human behavior in a complex economic environment.


Reid and Harrigan \cite{RN2689} conducted the first study on the entire cryptocurrency transaction history (up to mid-2011) and revealed emerging structure from in the Bitcoin flow network and demonstrated the transaction history's forensic capabilities. Since then, the data mining from cryptocurrency transactions has grown into a large body of research and been successfully applied in assisting multiple law enforcement actions, including ceasing the then-largest darknet market Silk Road in 2013 \cite{RN2311} and arresting suspects in a major theft from the then-largest cryptocurrency exchange Mt. Gox in 2017 \cite{RN3439}. To date, cryptocurrency transaction analysis, also called blockchain analysis, has become an essential means in fighting drug trafficking, computer network hacking, money laundering, and terrorism financing, as well as studying many other social-economical scenarios\cite{RN1257,RN568,RN1579}.

In this survey, we collect articles by searching in the Web of Science database using specific terms including ``cryptocurrency'', ``transaction'', ``Bitcoin'', ``Ethereum'', and several notable altcoins and payment protocols such as ``Monero'', ``Zcash'', and ``Lightning Network''. Search results appeared in many primary computer science and engineering \cite{RN2359,RN2578,RN3215}, physics \cite{RN2336}, and economic venues \cite{RN3155}. We focus our survey on the knowledge discovery, i.e., data mining of user behaviors, in cryptocurrency transaction records and identify three lines of work, namely: 
\begin{itemize}
    \item \textit{traceability and linkability analysis}, which addresses whether chains of transfers can be traced in the transaction records and whether blockchain addresses can be associated, i.e., linked, to the same identity;
    \item \textit{collective transaction pattern analysis}, which addresses the emerging user behavior by examining the cryptocurrency economy from a macroscopic perspective; and
    \item \textit{individual user behavior analysis}, which focuses on discovering transaction patterns from a single user or a particular type of user and using the patterns downstream machine learning tasks. 
\end{itemize}
We will present the background information for each line of work, summarize the methodologies, and discuss major findings in these works.

This survey contributes to the literature by providing a complete spectrum of knowledge discovery from cryptocurrency transactions and also serves as a handbook for researchers and practitioners interested in harnessing the concurrency transaction data in their own research. Nonetheless, we also recognize existing reviews that address individual subtopics covered in our survey, as follows. Technical introductions to current cryptocurrencies \cite{RN1758} and blockchain \cite{RN3367,RN3225} designs cover issues such as data models, consensus protocols of distributed ledgers, and system throughputs. \cite{RN1971,RN908,RN909} address the anonymity, privacy, and security issues of Bitcoin and Bitcoin-like cryptocurrencies. De-anonymization techniques using transaction records, mixing services, and designs of altcoins were surveyed and discusses. A survey of tools for smart contract code analysis can be found in \cite{RN3325}. Apart from the cryptocurrency transaction records, blockchain networks also produce other forms of data worthy of analysis. For example, traffic analysis of the blockchain peer-to-peer networks is particularly useful in understanding the systems' communication overhead and revealing user identities \cite{RN419}.

This survey is organized as follows. Section \ref{sec:preliminaries} provides basic concepts of the cryptocurrency economy and blockchain data models. Section \ref{sec:traceability} addresses the traceability and linkability analysis in Bitcoin, altcoins, and mixing service scenarios. Section \ref{sec:collective} addresses the collective patterns in cryptocurrency user behaviors, with an emphasis on emerging structural patterns in the transaction networks. Section \ref{sec:individual} addresses the transaction pattern analyses for a single user or a particular type of user. Section \ref{sec:tools} summarizes the extract, transform, and load (ETL); analytic; and visualization tools for cryptocurrency transactions as well as online intelligence platforms. Section \ref{sec:openproblems} discusses several open problems in the field. Section \ref{sec:conclusion} concludes.

\section{Preliminaries}
\label{sec:preliminaries}

\subsection{The cryptocurrency economy}
Early cryptocurrencies, such as Bitcoin and its derivatives, were merely used as payment media. Modern users also treat cryptocurrencies as investment or speculation targets in primary and secondary markets, or as tokens in gambling and recreational games.

\subsubsection*{Cryptocurrency as payment medium}
\textit{Mining rewards}. Mining, or minting, is the process of coin generation in proof-of-work (PoW)-like blockchain systems. The generation of cryptocurrencies requires solving a computationally heavy problem. The party who successfully solves a problem can get a certain amount of cryptocurrency as a payment for their resources spent. The mining process can be either an individual or a collective effort. Individual miners can contribute their computing resources to a mining pool and get rewards any time a peer miner solves the problem.

\textit{Faucet}. Like mining, cryptocurrencies can also be rewarded to users who complete generic tasks, such as solving a captcha. In this case, the party to disseminate cryptocurrencies is a human envoy. This rewarding process is also called a faucet.

\textit{Purchasing}. Laszlo Hanyecz made the first documented offline purchase with Bitcoin---10 000 BTC for two pizzas---back in 2010 \cite{website:LaszloHanyecz}. Nowadays, end-users of Bitcoin can use cryptocurrencies to make various purchases via online marketplaces or offline shops, e.g., multimedia content, electronics, and clothes. However, the most common purchases made using cryptocurrencies are drugs in darknet markets.

\textit{Wallets}. Online wallets are similar to banks, providing storage services for users' cryptocurrencies. Clients deposit to the wallet by transferring their cryptocurrencies to the wallet services' blockchain addresses and can make payments by sending from online wallet's addresses directly. Instead of maintaining their offline public/private key pairs, users use a pair of traditional username and password to access their wallets.

\textit{Ransom}. Attributing to the anonymous feature of cryptocurrencies, extorsions such as computer malware, human kidnapping, sextortion, and blackmailing often ask cryptocurrencies for ransom. Similarly, phishing emails asking for cryptocurrency transactions are also commonly seen.

\textit{Money laundering}. The fast-moving nature of cryptocurrencies provide an ideal channel for money laundering. Users can conceal the origin of illegally obtained money by buying cryptocurrencies and later selling them to make the money ``clean.'' Similar activities, such as bribing, also take advantage of the anonymity of cryptocurrencies. However, this process leaves traces in the cryptocurrency transactions and risks being exposed through transaction analysis.

\subsubsection*{Cryptocurrency as virtual assets}
\textit{Primary market}. Start-up projects and companies can issue their own cryptocurrencies to represent the equities of their projects. They sell their cryptocurrency for fiat money or other value-bearing media through crowdfunding activities, such as Initial Coin Offering (ICO) and Initial Equity Offering (IEO). Investors who buy these cryptocurrencies usually sell them in cryptocurrency exchanges for a profit later. Start-up projects can also use airdropping, i.e., sending out tokens to investors for free, to gain awareness of their projects.

\textit{Secondary market}. Bitcoin was first publicly traded in online cryptocurrency exchanges in 2010. Now, people can buy and sell cryptocurrencies with/for fiat money, cryptocurrencies, or other value-bearing media on these platforms. Exchanges can either execute users' selling and buying orders automatically or allow users to list and match their orders in a forum-like platform, i.e., in an over-the-counter (OTC) fashion. Exchanges also provide various cryptocurrency related financial products, such as futures and options. Some exchanges even conduct ``pump and dump'' schemes to manipulate with cryptocurrency prices.

\textit{Ponzi scheme and pyramid sellings}. Fraudulent activities are widely seen in the cryptocurrency economy. Ponzi schemes, or HYIP, are the most common types of fraud \cite{SEC}. A Ponzi scheme lures investors to its program by awarding early investors with an unreasonably high yield. However, when a new investment slows down, or the scheme organizers see fit, they stop giving out yields and take away all the investments. Ponzi scheme often involves pyramid selling, i.e., by allowing investors to sell the cryptocurrencies to their peer investors for profit.

\subsubsection*{Cryptocurrency as token}
\textit{Gambling}. Gambling games, such as dice games and roulette, use cryptocurrencies as chips. Gambling games are often the most active applications on blockchain networks.

\textit{Multiplayer games}. The introduction of the smart contract further enables better flexibility to game design. Games such as Cryptokitties allow users to generate tokens with random features, list them for sale, and buy or rent tokens from other users.

\subsection{Data Models}
\label{sec:datamodel}
Bitcoin and Ethereum are the two milestones, i.e., blockchain 1.0 and 2.0, in cryptocurrency and blockchain design. Most cryptocurrencies adopted and modified upon their transaction record data models \cite{RN3225}. Bitcoin and its derivatives use the unspent transaction output (UTXO)-based data models, while Ethereum and its derivatives use the account-based data models.

\subsubsection*{UTXO-based data model}

\begin{figure}
\centering
\includegraphics{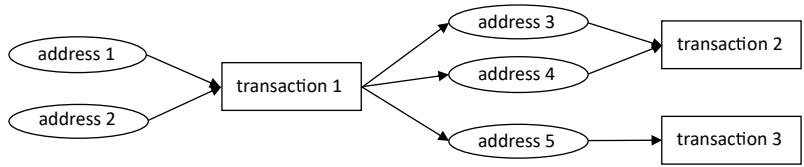}
\caption{UTXO-based transaction data model.}
\label{fig_utxotxn}
\end{figure}

In a typical unspent transaction output (UTXO) data model (see Figure \ref{fig_utxotxn}), addresses are the basic identities that hold virtual values. An address can be generated offline using Bitcoin's customized hash function to a public key generated by the Elliptic Curve Digital Signature Algorithm (ECDSA) to a user-specified random number as the private key. The transfers of values are settled by recording a transaction in the blockchain. In a typical UTXO transaction, e.g., transaction 1 in in Figure \ref{fig_utxotxn}, all the values stored in the input addresses, e.g., addresses 1 and 2, are transferred to output addresses 3, 4, and 5, with specific value allocations. The output addresses can be further used as the input addresses in the following transactions. Note that, theoretically, the maximum numbers of input and output addresses in a transaction are not limited, but since the size of a transaction record cannot be larger than the block size, the practical total number of input and output addresses has a limit.

Altcoins stand for alternatives for Bitcoin.  For example, Litecoin and Dogecoin are typical early altcoins that replicate most of Bitcoin's technical designs. Later altcoins, such as Zerocash and Monero, also adopt the UTXO-data model but use extra cryptography techniques to enhance their anonymity. Each altcoin has its own running blockchain network, which stores the transactions of this particular altcoin exclusively.

\subsubsection*{Account-based data model}
Blockchains can store not only transactions but also other formats of data, including text, image, and even computer codes. The code stored and executed in the blockchain database is also called a (smart) contract. Ethereum and its derivatives, such as Neo and EOS, use account-based transaction data models \cite{RN3435}, where the accounts are still blockchain addresses but can be either an externally owned account (EOA) or a contract account (CoA). While an EOA can be created using a similar method as an address in the Bitcoin blockchain, CoAs must be created by a transaction: an EOA or a CoA  sends a transaction to a ``null'' receiver with computer codes written in the auxiliary information. An address will then be generated by the blockchain system and assigned to the CoA. Both types of accounts can hold the blockchain's original cryptocurrency, and their current holdings are stored in the blockchain's running memory.

\begin{figure}
\centering
\includegraphics{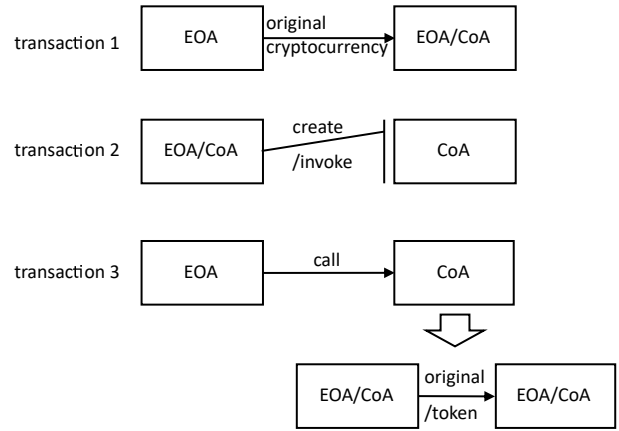}
\caption{Account-based transaction data model.}
\label{fig_account_based_txn}
\end{figure}

There are three possible types of transactions in an account-based data model. First, a transfer of the blockchain's original cryptocurrency from an EOA to an EOA or a CoA (transaction 1 in Fig. \ref{fig_account_based_txn}). Second, the creation of a contract or an invocation of the computer codes stored in a contract by transferring a zero-value original cryptocurrency to it with auxiliary information indicating the target function and a set of parameters (transaction 2 in Fig. \ref{fig_account_based_txn}). Third, a token transfer. Contract creators can install a virtual token in a smart contract, allow transfers of the token or part of it between blockchain identities through specific functions like \texttt{transfer()} or \texttt{distribute()} (transaction 3 in Fig. \ref{fig_account_based_txn}). The contract then records the changes to account balances. Note that in this case, only the amount of the original cryptocurrency or tokens held by EOAs or CoAs will be changed, but no actual transaction is explicitly logged in the blockchain.

\subsubsection*{Other transaction data sources}
Except for the transactions recorded in individual blockchain networks, transfers of cryptocurrency can also happen across different blockchains or even beyond blockchains.

The most popular blockchains, e.g., Bitcoin and Ethereum, occupy hundreds of gigabytes of space in computer storage and are ever-expanding on a daily basis. New technologies such as lightning network, sharding, and cross-chain transactions have been proposed to ease the management overhead and reduce the resources required to maintain the blockchain database. A lightning network enables users to create ``payment channels'' and conduct transactions in the channels before reporting the clearings to the blockchain. Sharding enables the blockchain network to store the entire transaction history but only requires each blockchain nodes to store a proportion of the transaction history. Cross-chain protocols enable users to swap cryptocurrencies across different blockchains. In these cases, the shattered or entangled transaction records add to the difficulty of transaction network analysis. 

Users trade cryptocurrencies with other users in exchanges. Most of the transactions in centralized exchanges are not publicly available, except for rare exceptions, such as leaked datasets from hacking of an exchange's database \cite{RN2335}. However, some OTC exchanges, e.g., bitcoin alpha \cite{RN1216} and bitcoin-otc \cite{RN2562}, on the contrary, disclose the transactions to the public. These transactions can be further related to blockchain transactions, providing an auxiliary information source of cryptocurrency flows.

\section{Traceability and Linkability Issues}
\label{sec:traceability}



\subsection{Tracing cryptocurrency transactions}
The transparency of cryptocurrency transactions enables forensic analyses of various crimes, using statistical analysis and graphical visualization techniques to the payment transactions of thefts, ransomware, sextortion, and illicit tradings. Several studies have demonstrated that forensic analyses can reveal the entire course of criminal transactions with surprisingly high accuracy. 

The most massive theft in cryptocurrency history was the Mt. Gox exchange hacking in 2011 \cite{RN3140}. The hackers allegedly stole more than 850 000 Bitcoin and led directly to the bankruptcy of the once largest cryptocurrency exchange in the world. As cryptocurrencies must be turned into fiat money for the thieves to profit, they must first transfer the stolen cryptocurrency into exchanges. Tracing analysis showed that the stolen bitcoins changed hands several times before landing in exchanges BTC-e, 0x, Bitcoinica, and CryptoXChange. Several interim addresses in the transaction flow, along with those from several other major thefts, passed through addresses belonged to Alexander Vinnik, the founder and main beneficiary of cryptocurrency exchange BTC-e. Vinnik was eventually arrested for alleged money laundering in 2017 \cite{RN3439}.

The Silk Road market was shut down by the Federal Bureau of Investigation (FBI) in October 2013. The FBI seized 114 336 Bitcoins, i.e., transferred them into an FBI-created arrest custody address. 89\% of the Bitcoins came from a set of 15 addresses at the end of a market escrow chain. However, among the distributions that peeled off the chain, more than 100 000 Bitcoins finally arrived into an address not relevant to the FBI \cite{RN2311}. It was believed that not all the darknet Bitcoins were seized, but some were detained by individual FBI agents who were later convicted for stealing in the Silk Road case \cite{RN3143}.

Cryptocurrencies are ideal ransom payment mediums owing to the anonymity feature. Many recent computer hacks, such as Cryptolocker and WannaCry, asked for Bitcoin for ransom. \cite{RN2345,RN2346} showed that payments to 35 families of ransomware, including the infamous CryptoLocker, CryptoWall, DMA Locker, and WannaCry, are all traceable, mostly to a handful of responsible parties. The economic impact of the ransomware from 2013 to mid-2017 was estimated to be USD 12 million minimum.

Moreover, cryptocurrency transactions can also reveal the monetary flows of human trafficking and sextortion. For example, \cite{RN1304} was able to uncover human traffickers by associating sex ads to specific Bitcoin transactions and addresses with 90\% accuracy. \cite{RN3011} tracked and investigated monetary flows of a series of sextortion campaigns and found that one single entity was likely controlling the financial backbone worth a minimum of USD 1.3 million.

\subsection{Counter-tracing measures}
In the light that the original design of Bitcoin's transactions is easy to trace, clever thieves and extortioners may use counter-tracing solutions to cover the trace of their activities. Typical solutions include mixing services provided by third parties and altcoins with intrinsic privacy-enhancing designs.

\subsubsection*{Mixing services}
\begin{figure}
\centering
\hspace*{-2cm}
\includegraphics{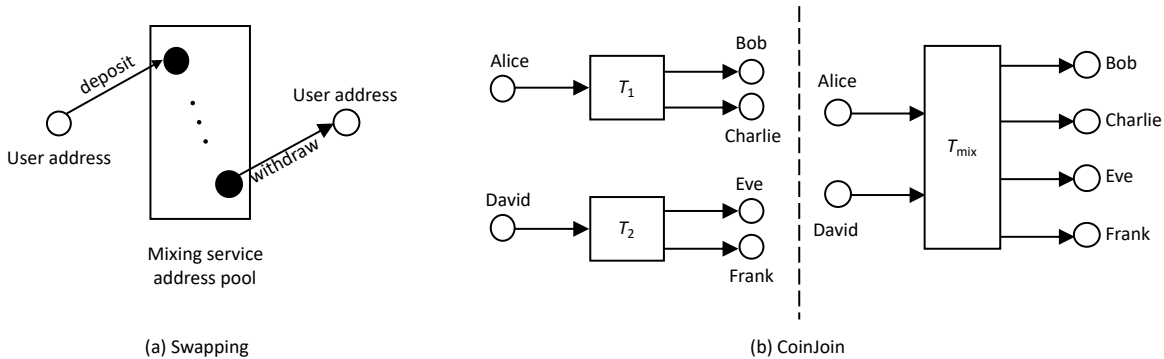}
\caption{Schematics of swapping and CoinJoin mixing.}
\label{fig_mixing}
\end{figure}

Mixing services aim to solve cryptocurrencies' traceability issues by merging irrelevant transactions. The two typical types of mixing methods are swapping and CoinJoin. A swapping-based mixing service (Fig. \ref{fig_mixing}a) accepts deposits from users to one of the addresses in an address pool and withdraw from another. Hence, the linkage between the deposit and withdraw addresses are disconnected. Mixing services using swapping include BitcoinFog, BitLaundry, and Helix. The CoinJoin (Fig. \ref{fig_mixing}b) mechanism allows two or more individual transactions (left) to be combined in a single CoinJoin transaction, which has the same presence as an ordinary multiple-input-multiple-output transaction (right) on the blockchain. Therefore, the relationship between real input-output pairs is obscured. CoinJoin-based services include JoinMarket, CoinShuffle, and Blockchain.info's SharedCoin (ceased service).

Mixing services are designed to hide the relationship between addresses in consecutive transactions. Therefore, they are often used for money laundering purposes for illicit activities. Under today's virtual asset market regulations, cryptocurrency exchanges can reject deposits from mixing service outflows under anti-money laundering (AML) regulation \cite{HKSFC}.

\subsubsection*{Altcoins with privacy-enhancing design}
This section covers the two major privacy-enhancing designs in altcoins, i.e., altcoins with decentralized mixing capabilities, such as Zerocoin, Zerocash, Zcash, and Dash; and CryptoNote design, including Monero, Bytecoin, and DigitalNote. We will use Zcash and Monero to illustrate the two lines of designs.

Zcash allows users to store and transact ZEC, i.e., the Zcash cryptocurrency, with two types of addresses (transparent and shielded) \cite{RN6425}. ``Transparent'' addresses transfer values to other addresses essentially the same way as Bitcoin, while ``shielded'' addresses make transactions in ``shielded pools''. In particular, when depositing into the pool, the recipient is specified using shielded addresses, i.e., \textit{z}-address, which hides the recipient but still reveals the sender, and withdrawing from the pool hides the sender but reveals the recipient. The cryptographical basis for the shielded pool is practical zero-knowledge proofs called \textit{zk}-SNARKs. From the perspective of data models, Zcash transactions resemble a swapping mixing pool pattern.

Though zero-knowledge proofs form virtual mixing pools, they suffer from the disadvantage of computational cost. CryptoNote-like cryptocurrencies, such as Monero, take another perspective, i.e., ring signature, to add complexity to the transaction records without causing much computational overhead \cite{RN6416}. A Monero transaction allows several outputs from previous transactions to be merged as its inputs, but only that some of the inputs can be ``decoy'' as their values are never transferred to the output.

\subsection{Taint analysis techniques}
Mixing services and altcoin designs add extra layers of complexity to the transaction record by hiding the association between blockchain addresses. Nevertheless, they can still leave sufficient information for ``taint analysis'', i.e., tracing monetary flow and linking related blockchain addresses. We categorize existing taint analysis techniques into several types of heuristics and the graph matching approach.

\subsubsection*{Heuristics}

\begin{figure}
    \centering
    \hspace*{-1em}
    \includegraphics{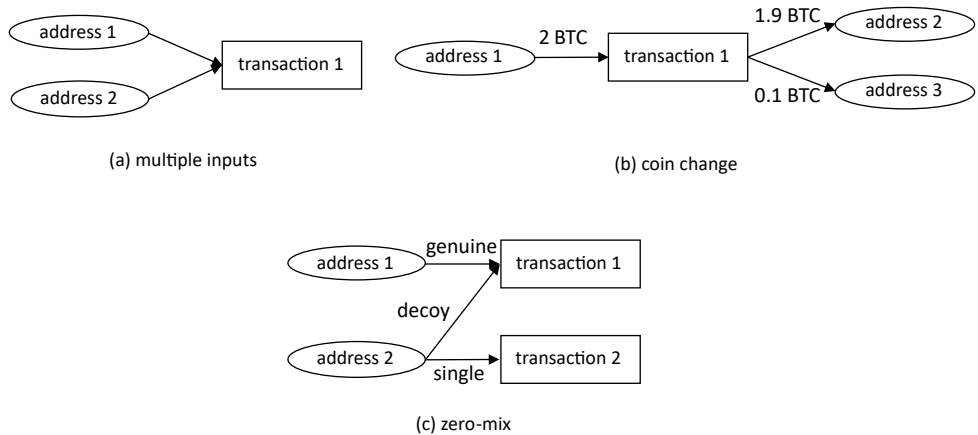}
    \caption{Commonly used heuristics in taint analysis: (a) multiple inputs, (b) coin change, and (c) zero-mix.}
    \label{fig:taint_graphheuristics}
\end{figure}

Commonly used heuristics in address associating and linking include the multiple inputs rule, the coin change rule, the zero-mix rule for cryptocurrencies with ring-signature design, and temporal heuristics. We formulate the first three types using graph representations, as shown in Figure \ref{fig:taint_graphheuristics}. 

\textit{Multiple inputs} is the most basic and widely adopted rule in associating UTXO addresses that potentially belong to the same user. When initiating a transaction, users have to sign the transaction with all of the input addresses' private keys. Therefore, all the input addresses in a transaction can be assumed to be owned by the same party \cite{RN2689,RN2359,RN2660}. For example, addresses 1 and 2 in Figure \ref{fig:taint_graphheuristics}a can be considered belonging to the same user.

\textit{Coin change}. The input value and output value in a UTXO transaction must be equal. If the input value is larger than the designated transaction value, the residue must be returned to a (usually new) address held by the transaction initiator, i.e., a change address. For example, addresses 3 in Figure \ref{fig:taint_graphheuristics}b is potentially a change in this transaction, and hence, belong to the same user who owned address 1.

The roll-out of the ``coin selection'' strategy in the official Bitcoin offline wallet, Bitcoin Core, in 2012 made the coin change rule even more obvious. When the user enters the amount of Bitcoin to be transferred to destination addresses, the client software automatically chooses the set of input addresses with an exact match to the value or a minimum change output. The change addresses usually hold only a small value and typically appear in transactions only once or twice. The coin change rule is usually used in conjunction with the multiple inputs rule \cite{RN2359,RN2660,RN2688}. 

Except for the most commonly used multiple inputs and coin change rules, other heuristics for Bitcoin transactions may also consider that all the output in a coinbase transaction belong to the same entity \cite{RN2693} or exploit specific transaction patterns, e.g., apparent self-transferring operations and those that resemble money laundering activities in a conventional banking system, to associate addresses \cite{RN2576}.

\textit{Zero-mix}, aka cascade attack or cascade effect, is a specific heuristic for CryptoNote cryptocurrencies. Let's assume the scenario as shown in Figure \ref{fig:taint_graphheuristics}c, where transaction 1 takes two unspent-inputs from addresses 1 and 2 at time \(t\) and transaction 2 takes the unspent-input from address 2 at time \(t+\delta t\). In this case, transaction 2 takes no foreign outputs used as mix-ins for the associated ring-signature and therefore is a zero-mix transaction. Hence, address 2 must be a real input in transaction 2 and a decoy in transaction 1 \cite{RN6416}.

Zero-mix in is a special case in the more general ``closed set'' attack proposed in \cite{RN2377}. If the number of inputs equals the number of distinct public-keys included in a CryptoNote transaction, i.e., forming a \textit{closed-set}, they are all real-spends in this transaction hence decoys in other transactions. \cite{RN2377} also proposed a computationally efficient realization of this attack. 

\textit{Temporal heuristics} exploits the timing of deposit and withdraw transactions in mixing services and altcoins transactions. For example, \cite{RN6408} found that temporally close deposit and withdraw transactions in Zcash's shielded pools have a high possibility of linkability. \cite{RN6416} also discovered that the ages of transaction outputs could be used to differentiate genuine and decoy inputs in CryptoNote transactions.

\subsubsection*{Graph matching}
Centralized and decentralized mixing services and altcoin designs follow a similar basis: merge irrelevant transfers into one blockchain transaction record to decrease these transactions' traceability. However, the fingerprints of transferred cryptocurrency values could still reveal the actual input-output pairs.

\begin{figure}
    \centering
    \includegraphics{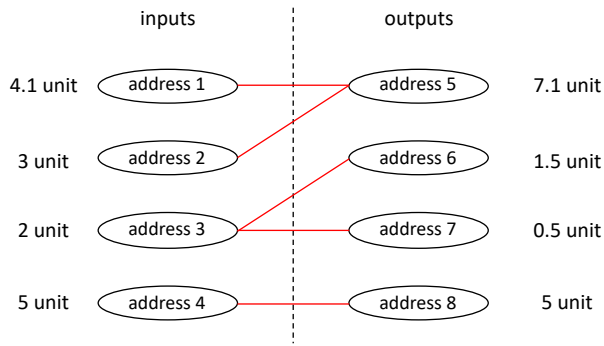}
    \caption{Illustration of the graph matching problem in tracing merged transactions.}
    \label{fig:graphmatching}
\end{figure}

Considering that users would deposit and withdraw the same amount of cryptocurrency to and from the mixing service, a widely adopted method to find input-output pairs is to find matched values or value combinations in the multiple-input-multiple-output transactions\cite{RN6416,RN1420,RN2123}. Given a set of inputs $I=\{I_1, I_2, ... \}$ and a set of outputs $O=\{O_1, O_2, ...\}$, the graph matching tries to find the exact match or the most probable matches between subsets of $I'\subseteq I$ and $O'\subseteq O$.  As shown in Figure \ref{fig:graphmatching}, the input values of address 1 and address 2 matches the output to address 5, and therefore, addresses 1, 2, and 5 might belong to the same user. Other matches in this transaction are also linked by the red lines. 

Note that graph matching is an NP-hard problem. Simplification of the problem can impose a time constraint, e.g., only consider temporally close input and outputs, to narrowing down the search for possible matches \cite{RN3132}. Special cases of the graph matching method are some times referred to as \textit{value fingerprinting heuristic} \cite{RN6425} or ``round-trip'' transactions \cite{RN6408}, where exact values match for single input and output in centralized or decentralized mixing pools. Similarly, matches of multiple digits between inputs and outputs \cite{RN3593} are also considered evidence of possible links between addresses.

\subsection{Findings}
Overall, Bitcoin and its replicates showed strong traceability: 87.6\% of transactions in Bitcoin, 88.5\% in Bitcoin Cash, and 85\% in Litecoin have a single output \cite{RN76}. As for linkability, using the multiple inputs rule, the number of clusters of associated Bitcoin addresses is approximately 45\%–70\% of the total number of addresses \cite{RN908}. When the coin change rule is applied with the multiple input rule, the number of associated Bitcoin address clusters further decreases to 22\%–37\% of the total number of addresses \cite{RN908}. Associating addresses can reveal addresses identities if one of the associated address is revealed. The largest cluster of associated Bitcoin addresses up to 2012 (156 722 addresses) were identified as the cryptocurrency exchange Mt. Gox, followed by well-known wallets and mining pools \cite{RN2313}. \cite{RN2329} successfully associated 968 unknown addresses to two addresses belonging to CryptoLocker found in Reddit.

However, heuristic rules are also prone to error. For example, the multiple inputs rule's effectiveness depends on repeated address use by a single user and can lose its efficacy if no UTXO address is reused or with CoinJoin-like mixing and ring signature altcoins \cite{RN2671}. Nick \cite{RN2660} tested the multiple inputs and coin change rules on a dataset containing 30 000 sets of Bitcoin addresses from the leaked BitcoinJ wallets, where each set of addresses belongs to an end-user. Using the multiple inputs rule alone, the associated clusters of addresses achieved a 68.6\% average recall ratio. With the coin change rule imposed, the average recall ratio only rose to 69.3\%, which accounts for less than 1\% of an increase in the accuracy. \cite{RN2659} showed that addresses clustered by mere heuristic rules could belong to several different entities, and therefore, proposed a probabilistic algorithm that utilizes known address tags to improve the heuristic rules.

Due to the lack of mixing services' ground truth data, reported precisions of mixing services' taint analysis are rare. Nonetheless, \cite{RN3132} used graph matching and temporal heuristics and found that 99.1\% of the input and output transactions performed by the mixing service provider Helix can be associated.

Although providing decentralized mixing functions, most Zcash activities are in the transparent part of the blockchain. Mining pools play essential roles in Zcash ecology. Mining rewards can be linked to 87.5\% addresses and 25.7\% transactions \cite{RN3526}, and that 95.5\% of the total number of Zcash transactions are potentially linkable to public addresses by just observing the mining activity \cite{RN3577}. What is more, 31.5\% of all coins sent to shielded addresses can be matched to public addresses using taint analysis techniques \cite{RN6408}. The primary users of the shielded pools are still mining pools: 65.6\% of the value withdrawn from the shielded pool can be linked back to deposits made by either founders or miners \cite{RN658}.


The multiple inputs rule loses its efficacy against CryptoNote designs, as the decoy addresses are usually not owned by the holders of the real inputs. However, careless usage of transaction outputs together in a new transaction can still reveal the ownership of the original transaction outputs \cite{RN6416}. Using zero-mix rule only, \cite{RN6415} was able to identify 62\% of the real inputs in Monero transactions with one or more mixes. The accuracy increased to 80\% when temporal heuristics are considered. \cite{RN2370} found that in 87\% of cases, the real Monero output from a redeemed can be easily identified with certainty with temporal heuristics. \cite{RN2377} were able to identify the real coin being spent in 70.52\% Monero inputs, 74.25\% Bytecoin inputs, and in 91.56\% DigitalNote inputs, using the zero-mix rule and their ``closed set'' attack.

\section{Collective Transaction Patterns}
\label{sec:collective}

Networks are ubiquitous in physical, technical, social, and economic systems with interconnected components. The study of complex networks in the past 20 years has shown that real-world networks possess universal underlying structural properties, such as scale-free property and small-world phenomena, and similar network formation mechanisms, e.g., preferential attachment \cite{RN3278}. Network analysis is widely adopted in the study of cryptocurrency transactions and has proven particularly useful in characterizing cryptocurrency user activities by examining these networks' temporal and structural properties.

\subsection{Transaction Network Construction}
\label{sec:networkconstruction}
A network $G=(V, E)$ consists of two sets of entities, i.e., a collection of nodes, $V$, and a collection of edges, $E$, connecting the nodes together. The construction of transaction networks for UTXO, account-based, and other transaction data models have subtle differences. This section summarizes three types of networks constructed from UTXO transactions, three types from account-based transactions, and discusses transaction network construction from the Lightning payment channels.

\begin{figure}
\centering
\hspace*{-1cm}
\includegraphics{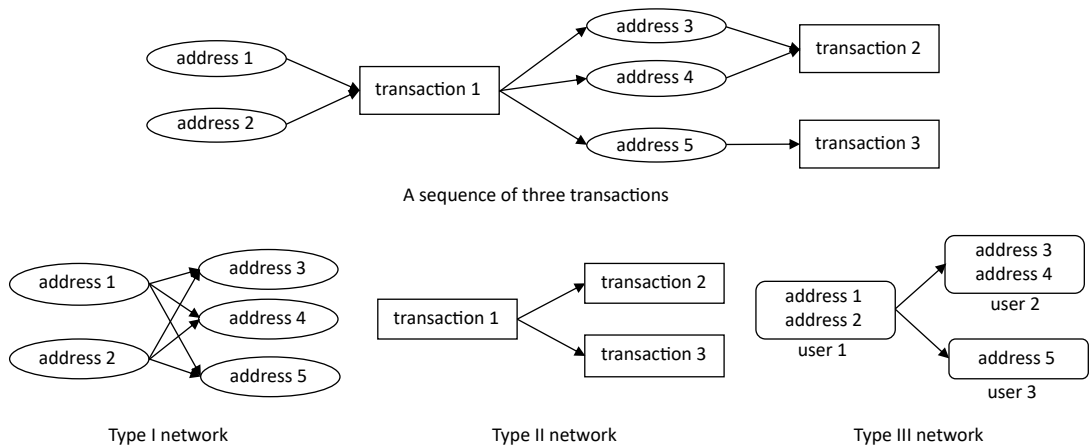}
\caption{Network construction from UTXO-based transactions.}
\label{fig_utxonet}
\end{figure}

\subsubsection*{UTXO transaction networks (Fig. \ref{fig_utxonet})}
 
\textit{Type I network}, or address network, uses addresses as nodes in the network and the flow of virtual values as edges \cite{RN3226,RN2655}. For example, input addresses 1 and 2 are connected to addresses 3, 4, and 5 with directed edges representing the value flow in transaction 1. The edges are usually unweighted because the actual transferred value between a specific pair of addresses is not explicitly given. Therefore, multiple-input-multiple-output transactions may create large cliques in the network.

\textit{Type II network}, or transaction network, uses transactions as nodes in the network, in which transactions are connected by the sharing of output/input addresses \cite{RN2123,RN2610}. For example, transaction 1 is connected to transaction 2 with directed edges, as the output addresses 3 and 4 of transaction 1 are used as the input addresses in transaction 2. The edges in this type of network can be weighted, i.e., the value being held in the interim addresses at that time. In some cases, Type I and Type II networks are combined, i.e., all the addresses and transactions are nodes in the network, while the input and output relationship between the addresses and transactions are considered edges \cite{RN3154,RN2540,RN2802}.

\textit{Type III network}, or user network, considers the flow of cryptocurrencies between users \cite{RN2313,RN2538}. However, since the blockchain addresses are anonymous, i.e., the addresses cannot be tied to user identities directly, association rules must be applied to associate addresses with potentially the same identity. Also, note that the Type II network is actually an interim state between the Type I and Type III networks in that it only considers all the inputs of a transaction being an entity but does not further merge all the entities with shared input addresses.

\subsubsection*{Account-based transaction networks}

\textit{Type I network}, or the original cryptocurrency transfer network, uses EOAs and CoAs as nodes in the network and the flow of non-zero original cryptocurrency as edges \cite{RN1765}. The edges are directed and weighted by the amount sent in the transactions.

\textit{Type II network}, or token transfer network, also uses EOAs and CoAs as nodes in the network but uses the flow of user-customized tokens as edges \cite{RN2554,RN2555}.

\textit{Type III network}, or invocation network, considers the creations and non-transactional function calls to smart contracts \cite{RN3215}. This network uses EOAs and CoAs as nodes, with directed edges pointing from the invoker addresses to the contract addresses, representing the creation or invoking relationship.

\subsubsection*{Lightning network}
Lightning Network (LN), launched in 2018, is the mainstream payment channel network (PCN). It attempts to relieve the pressure of ever-expanding Bitcoin ledger by creating payment channels across which any two users could exchange off-chain payments without burdening the entire network. The LN protocol specifies how to open payment channels and to route off-chain payments across a network of payment channels. A payment channel network is a network of payment channels where off-chain payments are routed \cite{RN2625}, rather than transactions between blockchain addresses. Information about live payment channels can be found on monitoring websites such as hashxp.org, 1ml.com, and lndexplorer.com \cite{RN3267}.


\subsection{Emerging structural properties}

\subsubsection*{Growth of the networks}

\begin{figure}
\centering
\includegraphics[width=4.5in]{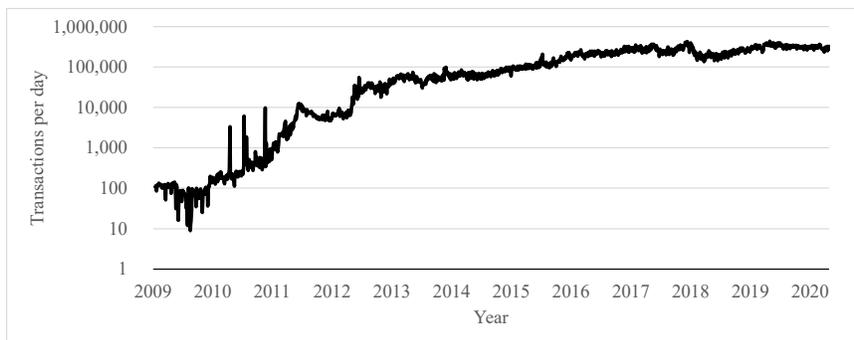}
\caption{Number of transactions per day in the Bitcoin blockchain network from 2009 to 2020.}
\label{fig_notxnpday}
\end{figure}

As of 2020, the total number of unique Bitcoin addresses appearing in the transactions was more than 600 million \cite{website:glassnode}, and that of Ethereum over 100 million \cite{website:etherscan}, with a daily increase of hundreds of thousands \cite{website:blockchain.info}. Therefore, new nodes and edges were introduced continuously into the transaction network, resulting in ever-expanding network size. As the first cryptocurrency, the growth of Bitcoin networks went through two stages: the initial stage (until the end of 2010) and the trading stage (since 2011) \cite{RN3807,RN6,RN220} (see Fig. \ref{fig_notxnpday}). The networks of other cryptocurrencies showed similar trends.

The growth rate of the networks depends on the adoption of cryptocurrencies. In the initial stage, Bitcoin was still an experimental idea used only by a small group of early adopters. The transaction volume was low, and the network structure fluctuated severely. The growth in the numbers of nodes and edges, network diameter, and the average distance between nodes correlated positively with the Bitcoin price \cite{RN2314,RN3160}. The rise of the Bitcoin price, in turn, attracted more attention to Bitcoin, which further expanded the user community and hence the network scale \cite{RN2703,RN2312}. The average balance of Bitcoin in the blockchain addresses decreased as the user community expanded \cite{RN2703}. In the trading stage, Bitcoin began to be accepted by a broader range of users and began to flow between addresses, resulting in a quickly expanding network size, and the structural properties, such as degree distribution and clustering coefficient, began to stabilize.

\subsubsection*{Centralized networks}
The degree distribution is the most common characterization of networks. Many real-world networks, such as the World Wide Web, movie actor collaboration network, and power grid networks, exhibit long tail, and sometimes power-law, distributed node degrees \cite{RN3429}. In these networks, most nodes have a limited number of neighbors, but some nodes can have a massive amount of connections. The degree distribution of the Bitcoin network (Types I and III) converged to a power-law distribution gradually over time, resulting in a scale-free network around 2010 \cite{RN1765,RN2314,RN1800}. The Type I  \cite{RN1765,RN1800} and Type III \cite{RN3469} Ethereum networks also exhibit power-law degree distributions with the power-law exponent $\gamma \approx2$.

Preferential attachment, one plausible mechanism driving real-world networks' evolution, refers to new nodes joining the network tend to connect to existing nodes with higher degrees. The preferential attachment was observed in Type I Bitcoin network's growth: hub nodes grow faster than low-degree nodes \cite{RN3807}. However, preferential attachment to the higher degree or richer nodes may not be an accurate mechanism to the Type III Bitcoin network's growth. \cite{RN220} proposed a fitness preferential attachment mechanism, where the fitness of a node $v$ is its potential to create new connections, i.e.,
\[{{f}_v(t)}=\frac{{{k}_v(t)}-{{k}_v(t-1)}}{\sum_{u=1}^{m}{({{k}_u(t)}}-{{k}_u(t-1))}},\]
where $t$ is the number of months starting from January 2019, $k$ is the degree of a node, and $m$ is the number of nodes in this month’s network. Different types of users have different intrinsic fitness: cryptocurrency exchanges are more attractive to connections than active traders, who are then more attractive than a common adopter of Bitcoin.

Another possible outcome of preferential attachment is that nodes with small degrees connect disproportionally to those with large degrees. This connection pattern is also referred to as disassortative mixing, e.g., nodes tending to connect to those with different structural properties. Most early-stage cryptocurrency transaction networks, such as the Type I Bitcoin network from 2009 to 2013 \cite{RN3807}, the Type I Litecoin network from 2009 to 2010 \cite{RN1800}, and the Type I Ethereum network from 2015 to 2017 \cite{RN2315} all showed disassortative mixing. In the cryptocurrency economy, disassortative mixing reveals that most transactions happen between end-users and popular services such as cryptocurrency exchanges, wallets, gaming, and gambling services, yet less frequently among the end-users themselves. However, the assortativity of the transaction networks increased over time. For example, the Type I Bitcoin and Type I Namecoin networks' assortativity eventually converged to 0 \cite{RN6}. This increment could have two causes: 1) high-degree nodes began to transfer cryptocurrencies among themselves, e.g., cryptocurrencies moving between multiple holding addresses with the same exchange, or 2) low-degree nodes could have started to have more interactions among themselves.

Not surprisingly, the Lightning Network also evolved into a centralized network. The early LN was disconnected, consisting of mostly small clusters and occasionally larger cliques \cite{RN488}. On June 13, 2018, the LN's snapshot contained 1355 nodes in the weakly connected component and 889 nodes in the disconnected periphery, exhibiting a degree distribution with $\gamma \approx2$ \cite{RN1134}. The LN gradually grew into a centralized network, with a distinct core-periphery structure. The Bitcoins distribution in each channel has an unequal Gini coefficient of the node strengths 0.88 in 2019 \cite{RN4559}. The centralized structure of the payment network exposes it under targeted attacks, i.e., a DDoS attack targeting hub nodes can remarkably sabotage the LN’s efficiency \cite{RN2738,RN4265}.

Since most commonly seen complex networks evolve power law in their degree distributions, a transaction network that deviates significantly from these rules is usually induced by anomalous activities \cite{RN3431}. Although the in-degree and out-degree distributions of the Bitcoin network generally followed the power-law distribution, some severe fluctuations exist in the distributions \cite{RN2538,RN2325}. \cite{RN2538} believed that these anomalies are caused by a deliberate transaction pattern called ``pseudo-spam chain,'' i.e., a large number of tiny value transactions. Similar disruptions can also be found in the transaction time interval distribution in the Ethereum blockchain: instead of following a strict power law, the distribution has some spikes at certain time intervals \cite{RN2565}. This anomaly was conjectured to be due to deliberate individual activities such as trading bots.

\subsubsection*{A small world}
Many complex networks show two characteristics when growing: densification, i.e., the increase in the number of edges in the network is super-linear to that of the nodes; and shrinking diameters, i.e., the average path length of the network shortens \cite{RN3283}. These characteristics are the indicators of small-world networks, along with a large clustering coefficient \cite{RN3281}.

Super-linearity between the increment of the number of edges, $M$, and the number of nodes, $N$, can be characterized by
\[M(t)\sim N{{(t)}^{\alpha }},\]
where $M(t)$ and $N(t)$ are the number of edges and nodes in the network at time $t$, respectively. If $\alpha>1$, super-linearity presents in the network evolution, and the average degree of all nodes will also increase over time. Super-linearity appeared in the early stages of transaction network growth. For the Type I Bitcoin network, the average out-degree of nodes increased from less than two in 2009 to around six in 2012 \cite{RN3226}. The nodes' out-degrees in the Type III Bitcoin network increased from around 2.6 in 2013 to around 3.1 in 2015 \cite{RN2538}. The Type I Ethereum network showed a general linear growth between the number of edges to the number of nodes, i.e., $\alpha$=1.0, from 2015 to 2017, but super-linearity emerged, i.e., $\alpha$=1.38 in the last 1/3 of the transaction record \cite{RN6}.

However, the super-linear growth of the transaction networks did not persist \cite{RN6}. Although $\alpha$=1.15 for the Type I Bitcoin network constructed from transactions spanning from 2009 to 2017, it decreased to 0.86, i.e., demonstrating a sub-linearity, between 2014 and 2017. The situation was similar in the Type I Namecoin network, in which $\alpha$ decreased from 1.05 to 0.99. A decreasing $\alpha$ means that users tend not to reuse previously used blockchain addresses and create new addresses in transactions to preserve their anonymity better. Therefore, the network evolution enters a new stage in which nodes' growth rate surpasses that of edges, and the network becomes sparse. However, the final stage may not happen to account-based blockchains such as Ethereum because the account creation procedure on these blockchains is much more complicated than UTXO blockchains. A less user-friendly address creation procedure may discourage users from creating new accounts on the blockchain.
 
``Six degrees of separation'' is a common metaphor for small-world networks, emphasizing a short average path length between each pair of nodes in the network. The average shortest path length in the Type III Bitcoin network’s largest connected component decreased from around five to around four from 2013 to 2015 \cite{RN2538}. On the contrary, the average shortest path length in Ethereum increased. The average distance between nodes in weekly snapshots of the combination of Type I and Type III Ethereum networks increased from around four in 2015 to five in 2018 \cite{RN2557}. The shrinking of the average distance in a Bitcoin network may be attributable to the introduction of popular services such as exchanges and wallets during that period. The expanse of network distance in the Ethereum network may be because a great number of new users were adopting the blockchain network and creating a huge number of new nodes, which were not yet densely connected.

A typical small-world network has a significantly larger clustering coefficient than a randomized network with the same size and density \cite{RN3281}. The average clustering coefficient of the Type III Bitcoin network was around 0.15 during 2011 and 2013 \cite{RN1765}. For the Type I Bitcoin network, the clustering coefficient was as high as 0.8 during 2009, but it decreased to 0.04 in 2014 \cite{RN1800} and stayed stationary around 0.05 afterward \cite{RN6}. The clustering coefficient of the combination of Type I and III Ethereum networks was close to 0 initially but increased to and remained at around 0.01 after block height 3 000 \cite{RN2557}. The clustering coefficients in different transaction networks were all higher than those in the randomized networks.

\subsubsection*{A bow-tie structure}

\begin{figure}
\centering
\includegraphics[width=3in]{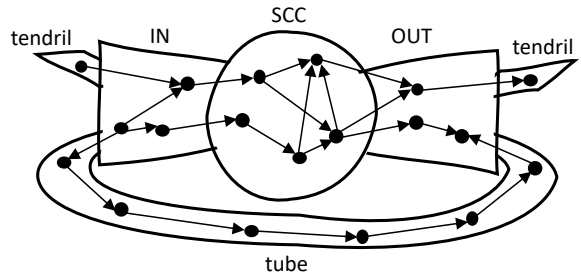}
\caption{A schematic illustration of the bow-tie structure in a directed network.}
\label{fig_bowtie}
\end{figure}

Large directed networks can evolve into a visually ``bow-tie''-like structure (Fig. \ref{fig_bowtie}). The bow-tie structure consists of four components: 1) a strongly connected component (SCC), which is the core of the network; 2) an in-component (IN); 3) an out-component (OUT), which are the sets of nodes reachable to and from the SCC, respectively; and 4) the tendrils, which are the sets of nodes unreachable to and from the SCC. \cite{RN2315} found that monthly snapshots of the Type I Ethereum network possess such a bow-tie structure. The SCC probably contains the hub nodes, e.g., exchange and wallets, and the IN and the OUT components are most likely the end-users. The tube component composed of tendrils can be regarded as extra bridges from the IN component to the OUT component.

\subsubsection*{Community structure}
A network contains community structure if it can be easily grouped into densely-connected sub-networks. Community detection algorithms are used as an enhancement to the heuristics rules of blockchain address association. For example, \cite{RN2908} conducted a two-step clustering process to Bitcoin addresses, i.e., first use the multiple inputs and coin change rules to associate addresses, then use Louvain algorithm to partition the transaction addresses into several communities further. They were able to find a set of CryptoLocker blackmail addresses using this process. \cite{RN3613} used multiple inputs and coin change heuristics to construct Type II Bitcoin network and used a community detection algorithm to partition the network into different activities further and improved the results obtained in \cite{RN2359}.

\subsection{Other collective patterns}
Preferential attachment not only results in a skewed degree distribution but also a centralized accumulation of \textit{wealth of blockchain addresses}, e.g., wealthier addresses accumulate cryptocurrencies significantly faster than the less wealthy ones \cite{RN3807}. The Bitcoin \cite{RN2313} and Ether \cite{RN1765,RN1800} wealth possession distribution in blockchain addresses both exhibit power-law. The highly wealthy addresses are not necessarily individual end-users but can also be exchanges or wallet services. Moreover, power-law distributions were also observed in the transaction value \cite{RN2314} and the time intervals between consecutive Bitcoin transactions initiated by the same addresses \cite{RN3807}.

The increasing centralization has also shown in the \textit{creation and usage of smart contracts}. Until 2018, smart contracts in Ethereum are only used to develop very simple token-centric applications, e.g., ICOs and crowdsales. Eighty percents of the smart contracts use at most 211 instructions \cite{RN3277}. Smart contract code similarity reveals substantial code reuse, where less than 10\% of user-created contracts are unique, and less than 1\% of contract-created contracts are so \cite{RN3060}. Moreover, contracts are three times more likely to be created by other contracts than they are by users \cite{RN3060}.

As a result, 0.05\% of the smart contracts are the target of 80\% of the transactions \cite{RN3277}, and that over 60\% of contracts have never been interacted with \cite{RN3060}. \cite{RN3052} surveyed 10 000 smart contracts source codes and a dataset of meta-data from etherscan.io. They found that the number of transactions and the balances of these contracts follow power-law and that the 20 smart contracts with the topmost number of transactions are all financial contracts.

\textit{Transaction fees} are the small amount of money that a user pays to the miners, i.e., the blockchain ledger keepers, when initiating a transaction. Depending on the busyness of the blockchain networks, fees may vary. The Bitcoin transaction fee per transaction surged at the end of 2017, to over USD 50, due to intensive network activities at that time and stabilized at several US dollars in 2020 \cite{website:blockchain.info}. It is notable that a non-negligible amount of Bitcoin addresses possess only a ``dust'' amount of values, i.e., they cost more in transaction fees to spend than the output value \cite{RN76}. Transaction fees in the Ethereum blockchain are called gas, which is the cost necessary to perform a transaction by miners. A transaction involving complex smart contract execution can have higher prices than an ordinary Ether transaction \cite{RN3816}. Miners can set the minimum gas price and decline to process a transaction if it does not meet their price threshold. \cite{RN3037} found that the number of pending transactions and the number of miners in the network have a significant influence on Ethereum gas fees.

\section{Individual Behavior Analysis}
\label{sec:individual}
An extensive literature focuses on characterizing and differentiating the activities of a specific type of agent in the cryptocurrency economy. The identities of blockchain addresses can be obtained from public online venues or interacting with known cryptocurrency services. Then, transaction features are extracted for the exploratory study of user behaviors as well as downstream machine learning tasks to classify the identities of blockchain addresses or look for anomalies in the transaction records.

\subsection{Tagging addresses}
Despite the anonymity nature of cryptocurrencies, service providers, such as exchange, wallets, and gaming, choose to disclose their blockchain addresses publicly. Some end-users also post their addresses in online marketplaces or forums to collect payments. Addresses related to major theft cases \cite{RN2689} and Ponzi schemes \cite{RN2384} can also be found on Reddit and BitcoinTalk. Researchers can also proactively collect addresses identities using cryptocurrency services and tracing the transactions from their own addresses \cite{RN2359}. Furthermore, the Ethereum Naming Service assigns humanly readable names to complex hash addresses, and therefore, can be used to reveal blockchain addresses' identities \cite{RN3280}.

The most commonly revealed addresses include cryptocurrency exchanges, merchants, escrow services, mining pools, gaming, gambling, and online wallets. Other less commonly found labels include mixing services, various scams including Ponzi schemes, ransoms, stolen bitcoins, and attackers. Start-up projects also disclose their holding addresses when disseminating tokens in the primary market. Note that, using the addresses association techniques mentioned in Section \ref{sec:traceability}, when one address in a node is tagged with a label, the label can be automatically inherited by all the other associated addresses.

Today's online intelligence platforms such as Blockchain.info \cite{website:blockchain.info}, Etherscan.io \cite{website:etherscan}, and WalletExplorer \cite{website:wallet} maintain lists of known blockchain addresses for user reference. Blockchain data analysis service providers, such as Elliptic \cite{website:elliptic} and Chainalysis \cite{website:chainalys}, also provide address labels to collaborators for research and law enforcement purposes \cite{RN2357}.

\subsection{Transaction features}
We categorize the commonly considered transaction patterns for individual addresses into four categories: volume, temporal, network structural, and contract code features.

\textit{Volume features} of an address include attributes such as the numbers of incoming/outgoing transactions, total volume of these transactions, balance, transaction fee paid, mining rewards revived, and aggregated figures such as the average, mean, and standard deviation of the previous features.

\textit{Temporal features} of an address include the activity period duration, activity intensity, and the average, mean, and variance of activity time intervals, as well as the skewness and kurtosis of the time interval distribution.

\begin{figure}
\centering
\includegraphics{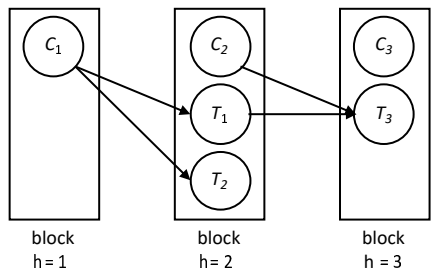}
\caption{Schematic of mint centrality.}
\label{fig_mintcent}
\end{figure}

\textit{Transaction network structural features} include node centrality, motif, network embedding, and neighbor identity information. Centrality is a measure of the importance of the nodes and edges in the network. Generic node centrality measurements include degree centrality, Eigenvector centrality, PageRank centrality, betweenness centrality, and closeness centrality \cite{RN3595}. Transaction network-specific centralities include the generalized entropic centrality proposed in \cite{RN681} and mint centrality proposed in \cite{RN2610}. The generalized entropic centrality of a node $u$ in Type III UTXO networks measures how likely a monetary flow goes from $u$ to any other node $w$. First, the probability of flow starting at $u$ and end with $w$ is defined by
\[
p_{uw} = \sum_{P\in\mathcal{P}{s,w}} \prod_{v\in P\\{s}} \tau_{P_v(v)} \frac{f(v',v)}{|\mathcal{S}(P_v)|},
\]
where $s$ is an auxiliary vertex that serves as
the source of the flow and has a single edge pointing to $i$; the sum is over all paths $P$ from $s$ to $w$; $\tau_{P_v(v)}$ is the probability that the flow goes out of $v$ on a particular subset of $\mathcal{N}(Pv)$, given the path $P_v$ was used to
arrive at $v$; and $f(v',v)$ is the amount of monetary flow from $v'$ to $v$. The mint centrality in Type II UTXO network of an address $A$ at a given block height $h$ is the number of distinct block heights of coinbase transactions with which address $A$ can be associated, through the transaction outputs it owned at any height prior to and including $h$. For example, in a Type II UTXO network as shown in Fig. \ref{fig_mintcent}, the squared boxes indicate different blocks; $C_{1}$ and $C_{2}$ are coinbase transactions; $T_{1}$, $T_{2}$, and $T_{3}$ are non-coinbase transactions. The mint centrality $mc(A,h)$ can be computed as follows:
\[mc(A,h)\text{=}\frac{1}{h}\sum\nolimits_{j=1}^{h}{rh}_{j},\] where $rh_{j}$  is either 1 or 0, representing whether or not the coinbase at height $j$ is linked to address $A$. Highly ranked Bitcoin addresses in mint centrality belong to SatoshiDICE as well as its associated addresses, faucet, and donation addresses.

Motifs are small building blocks that center around, start from, or end with a target node. In some networks, particular motifs appear more frequently than in the randomized network. These signature motifs in the network can often reveal the functional features of the real systems \cite{RN3269}. The smallest motif is a loop of two nodes connected by a pair of directed edges. A larger motif can consist of a particular connection pattern of three or four nodes. The clustering coefficient can also be categorized in this feature class because it calculates the number of triangles around a central node.

Network embedding encodes the structural features of a node into a low-dimensional space. Embeddings are usually constructed by network representation learning, i.e., an end-to-end training method that automatically transforms a network structure into a low-dimensional space. Early network representation methods include deepwalk \cite{RN3596}, node2vec \cite{RN3597}, or other customized biased random walks \cite{RN2546}. These methods capture the similarity between nodes as the overlap of neighbor nodes found by a random walk. By contrast, graph neural networks (GNN) take a network as the input and predefined node labels as the output and learn the nodes' feature vectors in an end-to-end learning scheme \cite{RN3598}. Since the transaction networks are temporal, i.e., the structure changes with time, temporal GNN models such as EvolveGCN are also used in transaction analysis \cite{RN2539}.

Neighbor identity features are usually dummy coded features indicating the existence of a labeled sample in a node’s neighbors.

\textit{Contract code features}. Source code and compiled code features are commonly used for the analysis of smart contracts. Code stylometry refers to the quantification and measurement of unique styles, e.g., wording frequency and the use of specific variable and function names \cite{RN3105,RN4555}. Furthermore, the symbolic analysis of programs' control flow can also yield multiple useful information for the identification of bugs or malicious contracts \cite{RN3048}.

\subsection{Signature behaviors of cryptocurrency economy agents}

\subsubsection*{Miners}
Blockchain networks reward the ledger keepers for their resources consumed. For example, a coinbase transaction is written as the first record in each block of the Bitcoin blockchain, containing a particular value of Bitcoin transferred from a ``no input'' address to one or more miner specified addresses. Bitcoin mining rewards were given to single miners in the early days when Bitcoin was only adopted by a small group of early players. However, when Bitcoin's price surged, the mining game changed. The difficulty of the mathematical problem that ledger keepers need to solve skyrocketed and single miners possessed little chance to solve a problem alone. Therefore, small miners formed or joined mining pools to pool their computational power and share the mining reward based on the resources invested \cite{RN3217}. \cite{RN4375} found that the percentage of pool-mined blocks was already 91.12\% in Bitcoin and 92.2\% in Ethereum, in July 2018. A handful of mining pools, such as F2Pool and AntPool, control most computational resources in the Bitcoin blockchain network \cite{RN3148}. Mining pool addresses are with the highest degree-based and betweenness centralities in the transaction networks \cite{RN2647}.

\begin{figure}
\centering
\hspace*{-2cm}
\includegraphics{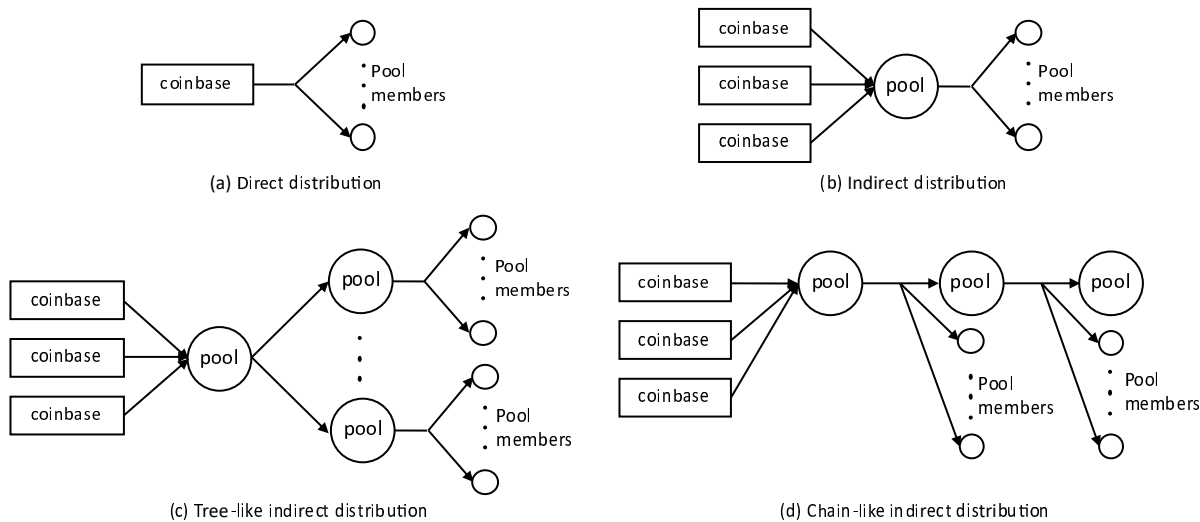}
\caption{Distribution modes of mining pools in PoW UTXO cryptocurrencies.}
\label{fig_miningdistribution}
\end{figure}

The transaction network structure helps to reveal the reward distributions among pool members in different mining pools. The simplest reward distribution method for a mining pool is to record a coinbase transaction with multiple output addresses, which directly belong to the miners (Fig. \ref{fig_miningdistribution}a). However, this distribution method has several downsides. First, when the mining pool grows big, e.g., to hundreds or more members, the mining reward cannot be distributed in a single transaction as the block size limits the maximum number of input and output addresses a transaction contains. Second, it requires a consensus of fixed distribution among members a priori to the actual mining task. Lazy miners could retain partial computational power in the actual task to receive a higher reward than deserved. Therefore, A natural improvement is that mining pools can use an interim reward holding address to aggregate all the output from coinbase transactions before distributing them among the pool members (Fig. \ref{fig_miningdistribution}b), e.g., F2Poll in 2014 \cite{RN1911}. Rewards to pool members can also be distributed in batches. For example, the mining pool ViaBTC divides the reward into multiple holding addresses and uses them to initiate multiple output transactions to the pool members (Fig. \ref{fig_miningdistribution}c). Another possibility is to initiate a multiple output transaction to the miners using the mining pool's holding address, but include another holding address in the outputs. Hence, the mining pool can further distribute mining rewards and eventually form a chain-like distribution pattern (Fig. \ref{fig_miningdistribution}d). Note that the output holding address in each chain transaction can be the input address (as used in BTC.com) or another new address (as used in AntPool) \cite{RN3146}.

The competition between mining pools became severe over the years. PoW-based coin mining is a process of looking for a random number whose hashed value falls into a specific range. Some mining pools simultaneously mined several blockchain ledgers with the same design, e.g., Bitcoin, Litecoin, Namecoin, Dogecoin, Huntercoin, and Myriadcoin \cite{RN1393}. However, such ``merged mining'' has operated at the edge of, and even beyond, the security guarantees offered by the underlying Nakamoto consensus for extended periods. Some other mining pools tried to exploit the cryptocurrency system design for a larger profit. For example, Ethereum blockchain not only rewards the winner of the mining mechanism but also rewards those who produced new but unused ledger updates (uncle blocks). \cite{RN3047} found that during May and July 2018, a swarm of 7 500 miners with conspicuously small hash rates orchestrated by a single adversary managed to receive 19\% of the total uncle block without competing directly for the primary mining rewards.

Miners might also hop from pool to pool to boost their reward \cite{RN3464}. An analysis of Kano and Slush mining pool members found that hopping miners' medium rewards were three times higher for those stuck to one pool \cite{RN3147}. However, no matter which mining pool they belonged, the miners all transferred the reward quickly into the same set of blockchain addresses, which belonged to exchanges, wallets, or gambling services \cite{RN3145}. The average interval between miners receiving the reward to the time of such transfers shortened from 138 days in 2009 to 1.5 days in 2013 \cite{RN1911}.

\subsubsection*{Darknet market escrow}
Darknet markets are online trading platforms hosted on covert computer servers, which are only accessible through encrypted networks. Illicit merchandise, such as drugs, weapons, and private data, are the most common commodities on darknet markets. Cryptocurrencies became the major payment medium of darknet markets, attributed to their anonymity nature. Payments between vendors and buyers have two modes. One is that the buyers transfer cryptocurrencies directly into the vendors' addresses. However, vendors and buyers have to make extra effort to build prior trust before making such transfers. In this case, an escrow service, in which a buyer first transfers cryptocurrency into an interim address owed by the market and the market later transfers the value on to the vendor once the trade is finished, is provided by the marketplaces.

\begin{figure}
\centering
\hspace*{-2em}
\includegraphics{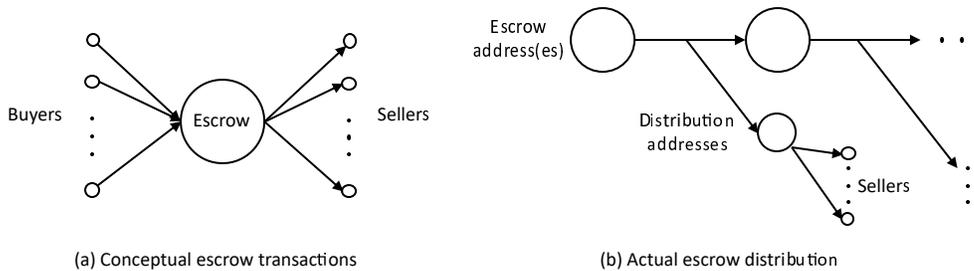}
\caption{Market escrow distribution pattern.}
\label{fig_escrow}
\end{figure}

Conceptually, buyers, sellers, and the escrow service form a star-like subgraph, with the escrow address in the middle and edges pointing from the buyers to the escrow and from the escrow to the sellers (Fig. \ref{fig_escrow}a). The escrow nodes naturally occupy the center of the transaction network. Popular darknet markets, such as SilkRoad, Agora, Wikispeed, and Evolution Market, ranked high in transaction volume, degree, Eigenvector, PageRank \cite{RN2655,RN2538} and betweenness centrality in the Type III Bitcoin network \cite{RN1765}. A high betweenness showed the diversity of darknet marketplaces' users because the market escrow nodes can connect nodes in different communities. Identifying an escrow address chain helps reveal market sale volume \cite{RN2679} or the sale of goods with particular prices \cite{RN2681}.

The tracing of escrow transactions has revealed the operation patterns of the market escrow services. Similar to the situation in mining reward distribution, the escrow address cannot transfer to an unlimited number of seller addresses in a single transaction. Therefore, a tree-like distribution or a chain-like distribution was observed from escrow transaction histories \cite{RN2359}. For example, A major chunk of the escrow is held in the escrow addresses, and payments to the vendors first arrive in distribution addresses by a smaller amount and are further forwarded to the sellers (see Fig. \ref{fig_escrow}b).

\subsubsection*{Mixing serivces}
Swapping-based mixing and CoinJoin-based mixing have distinct features in the transaction records. Swapping services have high daily transaction numbers \cite{RN2715} and transaction volumes \cite{RN3127}, short active periods, and low balances \cite{RN3134}. In the short time of activation, swapping addresses tends to act as a transit node, i.e., the middle node of a directed path of length 2, and rarely receives more than one transaction from the same address \cite{RN3134}. Some swapping service providers, such as Darklaunder, repeatedly use a single receiving address in a short time \cite{RN1375}, resulting in a larger degree of centrality than normal nodes \cite{RN3127}. Moreover, since users who belong to different communities may use the same mixing service, mixing service nodes can also act as bridges to nodes with few connections before mixings \cite{RN3632,RN2704}.

By contrast, CoinJoin-based transactions resemble ordinary multiple-input-multiple-output transactions. The average number of input addresses of Blockch-ain.info's SharedCoin mixing service transactions was 14.5 (between 4 and 40), and the average number of output addresses was 25.8 (between 4 and 42) \cite{RN2709}. Structurally, these transactions can be well disguised among normal transactions and lower the precision of address association rules in de-anonymizing Bitcoin addresses. \cite{RN3276} estimated that the volume of CoinJoin transactions was about 2.5\% of all Bitcoin transactions in June 2016.

A mixing service can also counter the taint analysis by making withdraw transaction values have a lower variant than normal transactions, i.e., creating similar outputs to prevent mixing re-identification \cite{RN3127,RN2708}. However, doing so also makes the mixing transactions more recognizable in the transaction network. \cite{RN2802} proposed an entropy measure to detect CoinJoin transactions in Bitcoin transaction history. For a multiple-input-multiple-output transaction $t$, its normalized input and output entropies
\[{{H}_{\text{in,norm}(t)}}=\frac{{{H}_\text{in}}(t)}{{{\log }_{2}}({k_\text{in}}(t))},\]
\[{{H}_{\text{out,norm}(t)}}=\frac{{{H}_\text{out}}(t)}{{{\log }_{2}}({k_\text{out}}(t))},\]
where $k(t)$ is the degree of $t$, and ${{H}_\text{in}}(t)$ is the absolute entropy of $t$, i.e.,
\[{{H}_\text{in}}(t)=-\sum\limits_{e\in {{E}_{t,\text{in}}}}{{{p}_{e,\text{in}}}{{\log }_{2}}{{p}_{e,\text{in}}}},\]
\[{{H}_\text{out}}(t)=-\sum\limits_{e\in {{E}_{t,\text{out}}}}{{{p}_{e,\text{out}}}{{\log }_{2}}{{p}_{e,\text{out}}}},\] 
where ${{p}_{e,\text{in}}}$ and ${{p}_{e,\text{out}}}$ are the values transferred in and out, respectively, through $e$, which is an incoming (outgoing) edge of $t$, normalized by the total number of Bitcoin inputs (outputs). A high entropy corresponds to a uniform distribution of bitcoin amounts in inputs (outputs) and indicates a potential mixing service.

\subsubsection*{Exchanges and wallets}
Since the launch of BitcoinMarket.com on March 17, 2010 (now defunct), cryptocurrency exchange activities occupied most of Bitcoin's blockchain space. The title of the largest exchange changed hands several times, e.g., Mt. Gox, Poloniex, and Binance. Overall, it was estimated that 18\% of the addresses belonged to exchanges in Bitcoin transactions from 2009 to 2015 \cite{RN2540}. Network nodes that belong to popular exchanges are highly ranked in degree \cite{RN1765,RN2314}, PageRank, betweenness, and closeness centralities in the transaction networks \cite{RN3215,RN2655,RN2647}.

Exchange nodes also have particular patterns of network connections. It was found that in the Type I Bitcoin network, the middle node of a directed path of length 2 that both starts and ends at an exchange address are highly likely to be another exchange address, indicating that many inter-exchange Bitcoin transactions occur \cite{RN2540}. However, those exchange addresses may not belong to the same exchange \cite{RN2654}.

Online wallets are like banks, accepting numerous deposits from users, and initiating a large amount of withdraws to user addresses. Wallet addresses ranked among the highest in degree, PageRank, and closeness centralities in transaction networks \cite{RN2655,RN1765}.

\subsubsection*{Games and Gamblings}
Blockchain technology perfectly suits the need for casino games with a transparent game logic, which can provide users with an extra source of trust in the games than in traditional online casino games. Therefore, casino games are the most popular applications on the blockchain. 

SatoshiDICE, a simple fortune redistribution game, was the most popular gambling game during 2012 and 2013, generating more than half of the transaction volume in the Bitcoin network \cite{RN2359}. Gambling transactions commonly have identical volumes, e.g., USD 1 or 0.01 BTC, and consecutive transactions between gambling services and players usually have short time intervals and high intensity \cite{RN2715,RN2661}. Transaction network nodes belonging to SatoshiDICE have a very high degree of centrality in both Type I and Type III Bitcoin networks \cite{RN2655,RN1765}. Other gambling services such as BTC Dice, BTCLucky, Clone Dice, and DiceOnCrac also have very high degrees of centrality in the Type III Bitcoin network \cite{RN2359}.

Smart contract-based games have gained popularity in recent years. As the most popular smart contract-based game, cryptokitties occupied caused a large scale Ethereum traffic congestion and raised the volume of pending transactions in the blockchain network from less than 1 000 to more than 10 000 at the end of 2017 \cite{RN6426}.

\subsubsection*{Market manipulation by whale addresses}
\label{sec:price}
The cryptocurrency secondary market was deemed to be highly inefficient. It is commonly believed that cryptocurrencies have no price baseline and that the volatility of the prices is mainly driven by the supply and demand of cryptocurrencies in the market \cite{RN3461}. Therefore, the change of majority in the numbers of buyers and sellers may lead to price changes; that is, if more cryptocurrencies are for sale, the prices drop, and vice versa \cite{RN3149}. \cite{RN3162} defined seller nodes in the Type III Bitcoin network as those who accept Bitcoin (and sell goods) and customer nodes as those who pay Bitcoin out. The customer to seller ratio has a correlation coefficient of 0.44 to the Bitcoin price. Empirically, cryptocurrency market prices were also found to be interactive with various assets, including gold \cite{RN2283}, the stock markets \cite{RN2503,RN2298,RN2300}, and among different cryptocurrencies \cite{RN28,RN2303,RN2168}. Prices were also found to be driven by social media \cite{RN2739,RN2350,RN84} and government regulations \cite{RN3020}.

It is believed that some users who have preliminary information in the cryptocurrency economy decide the market’s future movement \cite{RN2502,RN2475}. \cite{RN4266} traced hundreds of large Bitcoins transactions between 2018 and 2019 and found a positive abnormal trading volume for the 15-minute window before these transactions. \cite{RN2336} constructed daily Type III Bitcoin networks among the long-living nodes and most active nodes during 2012 and 2013; they found that the singular vector weights of Principal Component Analysis (PCA) analysis can explain the fluctuation of Bitcoin prices to a large extent, e.g., 0.85 Pearson's correlation. \cite{RN3155} constructed a combination of Type I and Type II Bitcoins, i.e., using both addresses and transactions as nodes in the transaction network. They considered each transaction a ``chainlet,'' which consisted of a central node, i.e., the transaction, and several incoming and outgoing edges connected to addresses. The number of particular chainlets, e.g., the number of input and output addresses $>20$, showed Bitcoin prices' predictive power.

A cryptocurrency exchange can also deliberately manipulate the supply and demand inside itself. The leaked Mt. Gox internal transaction records showed that one account bought USD 112 million worth of Bitcoins in a short 60-day window during September and November 2013 and caused the Bitcoin price to surge from lower than USD 200 to higher than USD 1 000 in two months \cite{RN2335}. The former managerial personnel later confirmed that the exchange operated this account. \cite{RN406} found that the transaction paths between the abnormal accounts that traded with a significantly higher or lower price than the market price formed many closed circles, i.e., self-loop, reciprocal edges, triangles, and polygons. These patterns resulted in a high face transaction volume, yet little actually changed hands. These patterns were also evidence of market manipulation. Among the cryptocurrencies, the ones with small market capitalization, a low traction volume, and trading in fewer exchanges were more prone to market manipulation \cite{RN3118}.

\subsubsection*{Ponzi scheme}
Early Ponzi schemes in the cryptocurrency economy used Bitcoin as an investment target. Later, the invention of smart contracts boosted the schemes. Since the HYIPs were written in codes and investment yields are automatically distributed to the investors, investors would establish blind trust in these seemingly transparent investment programs and ignore the possibility of dealing with a Ponzi scheme \cite{RN2327}.

Ponzi scheme addresses typically have lower transaction values, e.g., between USD 0.01 and 0.1; higher transaction frequencies than normal addresses \cite{RN17,RN2715}; and shorter average time span (37 days, as of 2014) than traditional offline Ponzi schemes \cite{RN2327,RN2996}. However, frequent interactions between the scheme organizers and end-users could prolong the active time \cite{RN2995}. The Gini coefficient of yield distribution and the proportion of incoming transactions could also help distinguish Ponzi scheme addresses from normal addresses \cite{RN2309,RN2327,RN2645}.


\subsection{Learning tasks using transaction features}
\label{subsec:mlidentity}

\subsubsection*{Address identity inference}
Address identity inference is a popular research topic in the cryptocurrency transaction analysis. This task, with transaction tracing, are also collectively known as the de-anonymization of addresses. With addresses tags broadly available, supervised machine learning algorithms are applied to capture the difference between labeled samples with the transaction features mentioned above.

Considering that the extracted features can be both numerical and categorical, the most commonly used classification algorithms are decision tree-based algorithms, including plain decision tree (DT), classification and regression trees (CART), gradient boosting-based decision trees (GBDT) (e.g., LightGBM, XGBoost), random forests (RF), and isolation forests (IF). Other commonly used algorithms include logistic regression (LR), naïve Bayes (NB), Bayes network (BN), supporter vector machine (SVM), and neural network-based methods, such as multilayer perceptron (MLP), long short-term memory (LSTM), convolutional neural networks (CNN). Learning algorithms that take the transaction network structure into account include graph convolution networks (GCN).

Identity inference tasks can be categorized into binary classification, e.g., mostly to identify whether an address is an illicit one (Table \ref{tab:binary}), and multi-class classification for generic types of econimic agents (Table \ref{tab:multiclass}). The most significant feature categories, best learning algorithms, and the best performances are also marked with a bold font. Categories with a lower identification accuracy are marked with bold font in the table.

\begin{sidewaystable}[htbp]
\centering
  \footnotesize{
  \caption{A summary of binary supervised learning tasks}
    \begin{tabular}{p{4em}p{6.6em}p{10em}p{20em}p{8em}p{10em}}
    \hline
    Research & Cryptocurrency & Target/Classes & Features & Best algorithm &  Performance
    \\\hline\hline
    \cite{RN2540} & Bitcoin & Exchange & Volume, \textbf{neighbor identity} & RF & 1.0 F1 \\\hline
    \cite{RN2309} & Bitcoin & Ponzi scheme & \textbf{Volume}, temporal & RF & 31 out of 32, 1\% FPR \\\hline
    \cite{RN17} & Bitcoin & Ponzi scheme & \textbf{Volume}, temporal, \textbf{motif} & RF & 95\% TPR, 5\% FPR  \\\hline
    \cite{RN2539} & Bitcoin & Illicit transactions & \textbf{Volume}, neighbor’s volume, embedding & EvolveGCN & 0.97 F1  \\\hline
    \cite{RN2850} & Bitcoin & Illicit transactions & \textbf{Volume} & RF & 0.98 F1  \\\hline
    \cite{RN3127} & Bitcoin & Laundering & Embedding (25 dimensions) & Adaboost & 0.94 F1  \\\hline
    \cite{RN3127} & Bitcoin & Laundering & Volume, temporal, neighbours, \textbf{embedding} & Adaboost & 0.93 F1  \\\hline
    \cite{RN2539} & Bitcoin & Laundering & Volume, temporal & RF & 0.76 F1  \\\hline
    \cite{RN3276} & Ethereum & Illicit & \textbf{Volume}, \textbf{temporal} & XGBoost & 0.99 AUC \\\hline
    \cite{RN2561} & Ethereum & Fraudulent & \textbf{Volume}, temporal & RF & 0.48 F1  \\\hline
    \cite{RN3281} & Ethereum & Ponzi scheme & Volume, temporal, contract code & J48 & 0.97 F1  \\\hline
    \cite{RN2439} & Ethereum & Ponzi scheme & contract code & RF & 0.79 F1  \\\hline
    \cite{RN3427} & Ethereum & Malicious addresses & \textbf{Volume}, temporal & RF & 47 out of 50 \\\hline
    \cite{RN2546} & Ethereum & Phishing scams & Embedding & SVM & 0.91 F1  \\\hline
    \end{tabular}%
    \begin{tablenotes}
    \item F1: F1 score; TPR: true positive rate; FPR: false positive rate; AUC: area under curve.
    \end{tablenotes}
  \label{tab:binary}%
  }
\end{sidewaystable}

\begin{sidewaystable}[htbp]
\centering
  \footnotesize{
  \caption{A summary of multi-class supervised learning tasks}
    \begin{tabular}{p{4em}p{6.6em}p{22em}p{12em}p{8em}p{10em}}
    \hline
    Research & Cryptocurrency & Target/Classes & Features & Best algorithm &  Performance
    \\\hline\hline
    \cite{RN2574} & Bitcoin & Exchange, gambling, service, and general & Embedding (8 dimensions) & DT & 0.91 F1 \\\hline
    \cite{RN3268} & Bitcoin & Exchange, service, gambling, mining Pool, mixer, and marketplace & volume, motif feature & cascading model & 99.7\% accuracy \\\hline
    \cite{RN4364} & Bitcoin & Mining pools, miners, mixing services, gambling, exchanges and others & \textbf{Volume}, temporal, embedding & RF & 0.96 F1 \\\hline
    \cite{RN2715} & Bitcoin & \textbf{Exchange/wallet}, faucet offering, \textbf{gambling}, \textbf{HYIP}, \textbf{marketplace escrow}, mining pool, mixer & \textbf{Volume}, temporal, motif & RF & 72\% accuracy \\\hline
    \cite{RN2579} & Bitcoin & \textbf{Exchange}, faucet, \textbf{gambling}, HYIP, \textbf{market}, mixer, mining pool & \textbf{Volume}, temporal, motif & GBDT, NN & 0.86 Macro F1 \\\hline
    \cite{RN2654} & Bitcoin & Exchange, service, gambling, \textbf{mining pool}, darknet marketplace & Volume, temporal, centrality, \textbf{motif} & GBDT & 0.91 F1 \\\hline
    \cite{RN2353} & Bitcoin & Exchange, hosted-wallet, personal-wallet, darknet marketplace, gambling, merchant-services, mining-pool, mixing, ransomware, scam, stolen-bitcoins, others & Volume, temporal, neighbor identity & Extra Trees & 96\% accuracy \\\hline
    \cite{RN4555} & Ethereum & 1071 smart contract authors & Stylometrics & RF & 91\% accuracy using source code, 80\% accuracy using byte code \\\hline
    \end{tabular}%
  \label{tab:multiclass}%
  }
\end{sidewaystable}

Binary classification tasks generally achieve very high accuracy, showing that illicit activities are highly separable in their transaction patterns. Comparatively, the accuracy of multiple classification tasks is significantly lower. In particular, exchange nodes are commonly confused with gambling, the marketplace, and Ponzi schemes in Bitcoin networks. Feature-wise, transaction volume features play essential roles in differentiating entities. Network structure-based features, especially certain motifs such as loops, also help improve the prediction power. Network embedded feature selection methods, when used alone, can also achieve reasonable performance compared to hand-picked features. As for learning algorithms, decision tree-based methods, especially random forests, achieved the highest performance in most of the tasks. GNN based methods, despite their novelty and popularity in recent literature, did not show superior prediction power to well-established methods.

The class imbalance problem is severe in the classification tasks, especially those involving illicit addresses. For example, fraudulent addresses are extremely difficult to obtain, while exchange and wallet addresses are abundant. Common solutions fall into two categories. Sampling-based methods, including the under-sampling of the large class \cite{RN17} and synthetic minority over-sampling (SMOTE) technique \cite{RN2353}, attempt to balance the number of samples in different classes. However, cost function-based methods can impose extra punishment to wrongly classified small class samples \cite{RN2309,RN2561}. Cross-validation and ensemble learning methods were applied to minimize the lack of samples in small classes.

\subsubsection*{Prices prediction}
Table \ref{tab:predictprice} summarizes the representative works of using transaction network features to predict cryptocurrency prices. Mean squared error (MSE), and root mean squared error (RMSE) are commonly used as precision indicators. The prediction results across different works may be incomparable as MSE and RMSE are sensitive to the cryptocurrency price scales. Nevertheless, it can be learned that network structural features, including both macroscopic properties, such as size and density, and microscopic properties, such as centrality measures of and motifs around nodes, can provide predictive power for the cryptocurrency price and price direction, e.g., achieving larger than 50\% accuracy, in the next time window. The predictive power of transaction network structural properties is also solid proof that the cryptocurrency market is highly inefficient.

\begin{sidewaystable}[htbp]
\footnotesize{
  \centering
  \caption{A summary of cryptocurrency prices prediction tasks}
    \begin{tabular}{p{4em}p{9em}p{9em}p{16em}p{10em}p{16em}}
    \hline
    \textbf{Research} & \textbf{Cryptocurrency} & \textbf{Target} & \textbf{Features} & \textbf{Methods} & \textbf{Best performance} \\\hline\hline
    \cite{RN3157} & Bitcoin & Price direction & Network size and density, mining difficulty, market information & \textbf{RF}, SVM, LR & 99\% accuracy in daily direction, 50\%–55\% accuracy in 10-minute direction \\\hline
    \cite{RN3149} & Bitcoin & Price and direction & Network size and density; volume features of the three most influential nodes & \textbf{LR}, SVM, NN, \textbf{Liner Regression} & MSE 1.94; 55\% accuracy in price change direction \\\hline
    \cite{RN3164} & Bitcoin & Price direction & Market information, \textbf{network size and density}, PageRank, and betweenness centralities & RF & Accuracy increased from 60\% to 74\% using centrality features \\\hline
    \cite{RN3344} & Bitcoin, Litecoin & Price and return & Motif features, market information & RF & Motif features contribute to the prediction of price and return \\\hline
    \cite{RN3158} & Bitcoin & Price direction & \textbf{Volume of the most frequently traded edges}, market information & NN & 60\% accuracy \\\hline
    \cite{RN3165} & Bitcoin & Price direction & Volume features, centrality features of the most influential nodes, market information &  \textbf{GBDT}, SVM, DT  & 82\% accuracy \\\hline
    \cite{RN3166} & Ethereum & Price & \textbf{Network size and density,} \textbf{distribution of node features}, market information & \textbf{LSTM}, CNN & Four times the error reduction with transaction network features than with only market information \\\hline
    \end{tabular}%
  \label{tab:predictprice}%
}
\end{sidewaystable}

\subsubsection*{Anomaly detection}
Bot activities, malicious attacks, and rare anomalous activities can also leave marks in the transaction records. However, the number of instances is too small to guarantee good classification performance in supervised learning tasks. Therefore, rule-based or unsupervised learning algorithms are used to identify outlier addresses in search for anomalous activities. The learning algorithms used are $k$-means, $kd$-tree, unsupervised SVM, Isolation Forest (IF), Gaussian Mixture Models (GMM), and a Role eXtraction (RolX) algorithm, which factorizes the feature matrix into two non-negative matrices and categorizes the nodes into clusters where smaller ones are considered anomalous \cite{RN3433}. A summary of transaction network anomaly detection methods and results can be found in Table \ref{tab:outlier}. Addresses with anomalous transaction patterns provide very strong predictive power to illicit activities such as the Ponzi schemes, thefts, and malicious attacks to the blockchain or cryptocurrency services.

\begin{sidewaystable}[htbp]
\footnotesize{
  \centering
  \caption{A summary of unsupervised learning-based anomaly detection}
    \begin{tabular}{p{4em}p{9em}p{9em}p{14em}p{21em}}
    \hline
    \textbf{Research} & \textbf{Cryptocurrency} & \textbf{Features} & \textbf{Learning algorithms} & \textbf{Outcome} \\\hline\hline
    \cite{RN1865} & Bitcoin & Volume, motif & $kd$-tree & Found 12 out of 30 known fraudsters \\\hline
    \cite{RN2325} & Bitcoin & Volume, motif & $k$-means, unsupervised SVM & Found 2 out of 30 known theft cases \\\hline
    \cite{RN2705} & Bitcoin & Volume & RolX  & Conjectured mixing service clusters \\\hline
    \cite{RN936} & Bitcoin & Network  & Visual inspection & Hidden communities about hundreds of users \\\hline
    \cite{RN1171} & Bitcoin & Volume, mining & PCA and visual analysis & Detected Pirate@40’s HYIP scheme \\\hline
    \cite{RN2270} & Bitcoin & Volume, temporal  & SVM & 15 out of 16 known attacks \\\hline
    \cite{RN3215}  & Ethereum & Volume, temporal & Threshold & Malicious attacks detected\\\hline
    \cite{RN6402}  & Ripple & Volume, temporal & IF, SVM, GMM & Bank fund movements, arbitrage bots, dumping\\\hline
    \end{tabular}%
  \label{tab:outlier}%
  }
\end{sidewaystable}

\section{Tools for Analyzing and Visualizing Transactions}
\label{sec:tools}
Part of the literature reviewed in the main text posted relevant codes or tools online. This section provides non-exhaustive lists of the analytics and visualization tools, as well as blockchain data analysis websites and services.

\subsection{ETL tools}
ETL tools extract information such as scripts, transactions, address balances, smart contract codes, and their current states from the blockchain, and feed the information into SQL, Graph, or NoSQL databases. ETL is the essential step before conducting a cryptocurrency transaction network analysis. Some tools ship with additional analytic features. Table \ref{tab:etltools} provides a list of ETL tools collected from the literature and online.

\subsection{Visualization tools}
Visual analysis is also a powerful tool in network analysis. Table \ref{tab:visuallist} summarizes the visualization tools that can construct transaction networks and provide visual analytic functions.

\subsection{Online intelligence platforms}
Online intelligence platforms are websites that provide in-depth blockchain information. Some platforms also allow users to post crowd-sourced knowledge to their databases. These intelligence platforms include Blockchain.info (now called Blockchain.com) \cite{website:blockchain.info}, Etherscan \cite{website:etherscan}, WalletExplorer \cite{website:wallet}, and BlockCypher \cite{website:blkcypher}. Technology companies such as Chainalysis \cite{website:chainalys} and Elliptic \cite{website:elliptic} also provide comprehensive services in cryptocurrency data analytics and malicious activity monitoring.

\begin{sidewaystable}[htbp]
\footnotesize{
  \centering
  \caption{List of blockchain data ETL tools}
    \begin{tabular}{p{14em}p{12em}p{6em}p{12em}p{20em}}
    \hline
    Tool & Cryptocurrency & Computer language & Database & Additional features \\\hline\hline
    BitIodine \cite{RN2361} & Bitcoin & Python & Neo4j & address clustering and classification \\\hline
    Blockchain2graph \cite{software:blockchain2graph} & Bitcoin & Java/typescript & Neo4j & Graph query with Cipher language \\\hline
    BlockSci \cite{RN3220} & Bitcoin, Litecoin, Namecoin, Zcash & C++/Python & in memory & MapReduce computation \\\hline
    BTCSpark \cite{RN3221} & Bitcoin & Cython & SQL & \\\hline
    bitcoin-blockchain-parser \cite{software:btcparser} & Bitcoin & Python & Local files & Supports SegWit \\\hline
    BlockETL \cite{software:blocketl} & Bitcoin & Java & SQL &  \\\hline
    Blockparser \cite{software:blkparser} & Bitcoin & C++ & in memory & Simple blockchain statistics: the closure of an address, all the block rewards and fees, taint analysis, etc. \\\hline
    rusty-blockparser \cite{software:rustyblkparser} & Bitcoin, Litecoin, Namecoin, Dogecoin, Myriadcoin, Unobtanium & Rust & csv files, MySQL & Simple blockchain statistics: average transactions per block, largest transactions, transaction types, etc. \\\hline
    Btctrackr \cite{RN2682} & Bitcoin & C++ & MySQL & Address clustering based on multiple inputs rule \\\hline
    BitcoinUses \cite{software:bitcoinuses} & Bitcoin & Java/Javascript & Hadoop & MapReduce computation \\\hline
    Blockchain-etl \cite{software:blockchainetl} & Bitcoin, Litecoin, Ethereum, Zcash, Dash, Dogecoin, Bitcoin Cash & Python & csv files, BigQuery &  \\\hline
    BlockchainVis \cite{RN3274} & Bitcoin & N/A & Accumulo & address clustering, mixing identification, and visualization \\\hline
    Ether\_sql \cite{software:ethersql} & Ethereum & Python & SQL &  \\\hline
    BlockAPI \cite{RN3224} & Bitcoin, Ethereum & Scala & MongoDB, MySQL, PostgreSQL, Fuseki & External data: exchange rates, address tags, protocol identifiers, etc. \\\hline
    EtherQL \cite{RN1527} & Ethereum & Java & MongoDB & Support range query and top-$k$ query \\\hline
    DataEther \cite{RN2172} & Ethereum & N/A & ElasticSearch & Ethereum account balance, transaction tracing, contract analysis \\\hline
    TokenScope \cite{RN3644} & Ethereum & N/A & N/A & Detecting inconsistent token behaviors with regard to ERC20 token standards \\\hline
    TEETHER \cite{RN660} & Ethereum & N/A & N/A & creating exploits for smart contracts \\\hline
    Erays \cite{RN3078} & Ethereum & N/A & N/A & reverse engineering tool for smart contracts \\\hline
    \end{tabular}%
  \label{tab:etltools}%
  }
\end{sidewaystable}%

\begin{table}[htbp]
\begin{adjustwidth}{-2em}{}
\footnotesize{
  \centering
  \caption{List of cryptocurrency transaction network visual analytic tools}
    \begin{tabular}{p{14em}p{8em}p{14em}p{4.5em}}
    \hline
    Tool & Cryptocurrency & Computer language & database  \\\hline\hline
    BitConeView \cite{RN3222} & Bitcoin & Python & N/A  \\\hline
    BitExTract \cite{RN3223} & Bitcoin & Python & MangoDB  \\\hline
    BiVA \cite{RN1303} & Bitcoin & Python & Neo4j \\\hline
    SuPoolVisor \cite{RN3681} & Bitcoin & interactive visual analytics and surveillance of mining pools & N/A \\\hline
    goBlockchainDataAnalysis \cite{software:goblk} & Faircoin & Go, NodeJS, AngularJS & MongoDB \\\hline
    \end{tabular}%
  \label{tab:visuallist}%
  }
\end{adjustwidth}
\end{table}%

\section{Open problems}
\label{sec:openproblems}
Despite the fruitful findings in the existing works, knowledge discovery studies from cryptocurrency transactions can advance in both methodology and the research questions.

\subsection*{Network representation learning}
To the best of the authors' knowledge, the cryptocurrency transaction networks are the largest networks that could be built from public data sources, with evolving network structures and abundant labels on the nodes and edges, providing much information for the knowledge discovery in transaction records. The state-of-the-art methodology in network analysis is network representation learning. Although this method has been used in recognizing user identities in the transaction networks, it is still out-performed by the traditional decision tree-based method in simple binary classification tasks \cite{RN2539}. Therefore, we believe that there is still considerable room for further development and application of network representation learning techniques in cryptocurrency transaction analysis.

\subsection*{Tracing across ledgers}
 From the cryptocurrency economic perspective, transactions across different ledgers are potentially intertwined, as users often change their holdings of one cryptocurrency to another. The linkage between different cryptocurrencies can be studied from several perspectives. For example, the transactions that happened inside cryptocurrency exchanges \cite{RN2379} and on exchange-like blockchain ledgers, such as Ripple \cite{RN6401}, can both be utilized further to trace the flow of monetary flow between users regardless of the actual currency used. Recently proposed cryptocurrency blockchain systems, e.g., Polkadot, also provide ``interoperability'' across different ledgers. In such a design, the transaction records of an amount of cryptocurrency can be ``locked'' on one blockchain and recorded on another. The study of multiple blockchain ledger integrations is just starting. There is no doubt that current methods can partially solve these questions, but many open questions remain, e.g., how to integrate multiple pieces of transaction histories, trace transactions among different cryptocurrencies, and detect illicit activities from these transactions.

\subsection*{New tokens standards}
More than 300 000 user-customized tokens have been issued on the Ethereum blockchain as of 2020, following the ERC20, ERC721, and ERC777 standards. ERC20 is the initial version of the token standard, ERC777 is an update to ERC20, and ERC721 is fundamentally different. ERC20 and ERC777 tokens are fungible, which means that each token can have a certain amount of distribution, and proportions of the whole amount can circulate among blockchain addresses independently. Comparatively, ERC721 tokens are designed to represent ownership over digital or physical assets. They are non-fungible, meaning that each token is of only one instance and cannot be further divided. Fungible and non-fungible tokens can have a distinct nature of circulation. What are the most distinct characteristics between the two types of networks? What can the distinction tell about user behaviors? Such questions are worth further investigation.

\subsection*{Primary market activities}
Some user-customized tokens issued on the blockchains are classified as securities by government regulatory bodies \cite{website:investor}. Companies and start-up projects raise funds by selling tokens to institutional and individual investors. This funding method, which bears the names of ICO, IEO, and security token offering (STO), is much like conventional primary market activities. However, because of the lack of regulation to this funding method in most parts of the world, fraud and scams frequently happen and often lead to investors' huge losses. A study of economy agent behavior in the funding process could help reveal the financial misconduct of token issuers and distributors, hence better protecting individual investors' interests.

\subsection*{De-Fi}
The development of smart contract-based financial instruments and the increasing adoption of blockchain technology in investors have brought a new wave of cryptocurrency innovation --- De-Fi. Short for decentralized financing, an umbrella term for the whole spectrum of financial activities over blockchain, De-Fi aims to disrupt and automate the entire financial industry. For example, users can deposit a digital asset into a smart contract as collateral for a loan of another digital asset. Other decentralized financial services, such as trading, lending, investment, asset management, and insurance, are also being developed. With all the transaction and usage history recorded on the blockchain ledgers, a thorough study of human financial behaviors based on an unprecedented rich dataset can be anticipated.

\subsection*{Digital fiat money}
The knowledge revealed by the transaction data stored in the blockchain networks not only helps to demonstrate the validity of the claim that the transparency of blockchain could facilitate auditing and regulating user activities, but could also be applied to a broader area beyond blockchain. Year 2020 sees the pilot tests and promotions of digital fiat money, such as Sweden's E-Krona and China's digital currency electronic payment (DC/EP). Although not necessarily adopting blockchain systems, the transaction records of these electronic versions of fiat money will be fully archived, e.g., in a central database. The successful experience of cryptocurrency transaction network analysis can be further borrowed by the electronic fiat money system in the future to help governments fight against bribing, money laundering, and terrorism financing.

\section{Conclusion}
\label{sec:conclusion}
Since the invention of Bitcoin in 2008, cryptocurrency has received wide acceptance among millions of users worldwide. A complete trace of users' activities and behaviors has been faithfully recorded on the blockchain. Having begun to notice the richness of the blockchain database a few years ago, academia has since produced a large body of research regarding cryptocurrency transactions.

The most extensively studied cryptocurrencies are Bitcoin, Ethereum, user-customized tokens issued on the Ethereum blockchain, and altcoins that provide extra privacy-preserving functions. Bitcoin is the representative cryptocurrency using a UTXO data model to store their transactions, while the Ethereum blockchain is the representative account-based data model.

Attributed to the transparency of blockchain ledgers, most of the cryptocurrency transactions are traceable and linkable. Although various coin mixing services and privacy-enhancing alternative cryptocurrencies are proposed, Careless use still can largely reveal user identity and behaviors. The tracing of illicit money flows between Ponzi scheme organizers and their victims, thefts, laundering, and ransomware victims have also provided strong evidence for solving these crimes.

From a macroscopic view, the cryptocurrency transactions among blockchain accounts form large and complex transaction networks. These networks are continually growing, with new blockchain addresses being created and used. Preferential attachment is the principal law governing the networks' growth: new nodes connect to existing nodes with higher connectivity. The networks eventually evolved into scale-free, e.g., with power-law degree distributions, and small-world networks, e.g., with short average path lengths and high clustering coefficients. The transaction networks can also show a bow-tie structure, with a large strongly connected component and obvious source and sink communities.

Agents in the cryptocurrency economy may have different economic behaviors and therefore form different transaction patterns. For example, mining pools send rewards to pool members in a tree-like or chain-like series of transactions. Marketplaces also use such distribution patterns to conduct their escrow services. Major agents such as cryptocurrency exchanges, online wallets, marketplaces, gambling games, and mixing services were all found to have the highest transaction volumes and possess transaction networks' central positions. These structural features can be further utilized in machine learning algorithms to derive models that differentiate and identify economic agents in the transaction network. 

With the cryptocurrency economy booming in recent years, we can foresee an abundance of new and disruptive innovations, especially blockchain-enabled financial services. We believe that cryptocurrency transactions will continuously provide new knowledge of various human social-economic behaviors in the future.

\section*{Acknowledgments}
The authors thank the valuable discussions from Prof. Ron G. Chen, Prof. Xiao-Ke Xu, and Prof. Ye Wu.

\bibliography{reference}

\end{document}